\documentclass[amsmath,amssymb,aps,pra,reprint,superscriptaddress]{revtex4-2}
\usepackage{graphicx}
\usepackage{color} 

\begin{document}
\title{Point-group symmetry analysis of many-electron wavefunctions on a quantum computer}

\author{Rei Sakuma}
\affiliation{Materials Informatics Initiative, RD Technology \& Digital Transformation Center, JSR Corporation,
  3-103-9 Tonomachi, Kawasaki-ku, Kawasaki, 210-0821, Japan}
\affiliation{Quantum Computing Center, Keio University, 3-14-1 Hiyoshi, Kohoku-ku, Yokohama 223-8522, Japan}

\author{Kenji Sugisaki}
\affiliation{Deloitte Tohmatsu LLC, 3-2-3 Marunouchi, Chiyoda-ku, Tokyo 100-8363, Japan}
\affiliation{Quantum Computing Center, Keio University, 3-14-1 Hiyoshi, Kohoku-ku, Yokohama 223-8522, Japan}
\affiliation{Centre for Quantum Engineering, Research and Education, TCG Centres for Research and Education in Science and Technology, Sector V, Salt Lake, Kolkata 700091, India}

\author{Shu Kanno}
\affiliation{Mitsubishi Chemical Corporation, Science \& Innovation Center, Yokohama, 227-8502, Japan}
\affiliation{Quantum Computing Center, Keio University, 3-14-1 Hiyoshi, Kohoku-ku, Yokohama 223-8522, Japan}

\author{Toshinari Itoko}
\affiliation{IBM Quantum, IBM Research--Tokyo, 19-21 Nihonbashi Hakozaki-cho, Chuo-ku, Tokyo 103-8510, Japan}
\affiliation{Quantum Computing Center, Keio University, 3-14-1 Hiyoshi, Kohoku-ku, Yokohama 223-8522, Japan}

\author{Hajime Nakamura}
\affiliation{Quantum Computing Center, Keio University, 3-14-1 Hiyoshi, Kohoku-ku, Yokohama 223-8522, Japan}

\date{\today}
\begin{abstract}
A point group is a set of spatial symmetry operations in molecular systems and is an indispensable tool
for analyzing molecular orbitals and spectroscopy experiments in chemistry.
Several quantum algorithms to exploit this symmetry have been proposed, but practical implementations
of point-group symmetry operations and the detailed symmetry analysis of
realistic many-electron wavefunctions are still missing. In this work,
we propose an ancilla-free hybrid method to analyze point-group symmetries of many-electron states,
which works for both abelian and non-abelian groups.
For a given wavefunction, our method calculates the projection weights of point-group irreducible representations
by applying orbital rotations derived from the eigenvectors of the representation matrices,
making it applicable to arbitrary basis functions.
The usefulness of our approach is demonstrated through numerical simulations of benzene and ferrocene molecules.
Furthermore, we perform a hardware demonstration of the weight calculation of the ground state
and the first excited state of benzene in $D_{2h}$ symmetry,
using up to 32 qubits of IBM's \texttt{ibm\_kawasaki} device. By combining
a tensor-network based encoding scheme
and error mitigation techniques,
we find the weights of irreducible representations for both states are faithfully reproduced within a few percent error.
Our results suggest that the proposed method serves as a practical tool for analyzing symmetry properties of many-electron wavefunctions in realistic material simulations
on near-term and early fault-tolerant quantum computers.
\end{abstract}

\maketitle


\section{introduction}
Symmetry is a fundamental concept in quantum mechanics, and quantum computing algorithms
significantly benefit from considering symmetries of a given system.
For example,
the conservation of the total electron number in the molecular electronic structure Hamiltonian,
related to $U(1)$ symmetry~\cite{PhysRevB.83.115125},
leads to a simplified circuit ansatz~\cite{Gard2020} or a reduction in the number of qubits~\cite{bravyi2017taperingqubitssimulatefermionic}.
Therefore, finding and utilizing symmetries of a given quantum state is
crucially important for realizing quantum simulations on near-term devices and early fault-tolerant quantum computers.

A point group~\cite{bishop} is a finite group of spatial symmetry operations in molecular systems and is
widely used in theoretical and experimental chemistry. It also plays key roles in understanding
intriguing physical phenomena and chemical reactions, such as the Jahn-Teller effect~\cite{10.1098/rspa.1937.0142,10.1098/rspa.1938.0008,bersuker}
and the Woodward-Hoffmann rules~\cite{Woodward1965}.
Quantum algorithms have also been proposed to exploit this symmetry~\cite{Setia2020,Picozzi_2023,PhysRevA.105.062452}.
One of the most useful applications of point group symmetry
is the symmetry-adapted projection of
many-electron wavefunctions~\cite{10.1063/1.5110682,PhysRevA.111.052433,khinevich2026symmetryadaptedstatepreparationquantum};
by projecting a general many-electron state onto 
a specific irreducible representation of the point group, this operation allows for obtaining
not only the ground state but also excited states of the system.
However, while the general methodology of symmetry-adapted projection has been discussed, a concrete implementation of point group symmetry operations
and the analysis of symmetry properties of
realistic many-electron wavefunctions remain largely underexplored.

In this work, we aim to bridge this gap by presenting a practical framework for
analyzing point group symmetry properties of realistic many-electron wavefunctions.
For this purpose, we focus on 
calculating the weights of irreducible representations in a given wavefunction
as a tool to analyze its symmetry properties.
We propose an ancilla-free, quantum-classical hybrid method to 
calculate this quantity, which
works for both abelian and non-abelian point groups.
This method leverages the fact that, under the Jordan-Wigner transformation,
point group symmetry operations
can be implemented on a quantum circuit with $O(1)$ depth
for commonly used symmetry-adapted molecular orbitals (e.g. Hartree-Fock orbitals)
and with $O(n)$ depth for general orbitals, where $n$ is the number of spatial orbitals.

The usefulness of this approach is showcased through numerical simulations 
of two prototype molecules, benzene and ferrocene.
To be specific, through a detailed analysis of single Slater determinants and local unitary cluster Jastrow functions~\cite{D3SC02516K},
we show that our method yields practical information about the point-group symmetries of many-electron wavefunctions,
and that it is also useful for assessing the quality of given trial wavefunctions.

Furthermore, as a proof of concept calculation,
we present a hardware demonstration of calculating the weights of irreducible representations
for benzene in $D_{2h}$ symmetry,
using up to 32 physical qubits of IBM's superconducting quantum device \texttt{ibm\_kawasaki}.
We prepare the ground-state and the first excited-state wavefunctions
of this molecule with
density matrix renormalization group (DMRG)~\cite{PhysRevLett.69.2863,SCHOLLWOCK201196},
and 
use a tensor-network based technique similar to Ref.~\cite{doi:10.1073/pnas.2425026122}
to classically convert them into a brick-wall ansatz.
We note
that for the DMRG wavefunctions expressed in a matrix
product state (MPS) form, the same calculation can be done
classically using matrix product operators (MPO)~\cite{SCHOLLWOCK201196}.
However, we emphasize that our approach can be applied
to arbitrary many-electron quantum states, including
strongly-correlated states or time-evolved states that
cannot be described efficiently in tensor-network based
methods.

To extract ideal noise-free results,
we compare two error mitigation techniques:
(i) zero noise extrapolation (ZNE)~\cite{PhysRevLett.119.180509}
with gate-folding implemented in Qiskit~\cite{javadiabhari2024quantumcomputingqiskit},
and (ii) quasi-probabilistic error mitigation approach
implemented in
QESEM~\cite{aharonov2025importanceerrormitigationquantum,aharonov2025reliablehighaccuracyerrormitigation}.
Our demonstration shows that the proposed approach
is useful not only for symmetry analysis of a given many-electron wavefunction, but also
for benchmarking real quantum devices and error mitigation techniques.

The rest of the paper is organized as follows. Section~\ref{sec:theory}
outlines point group symmetry and explains our proposed algorithm.
Section~\ref{sec:num} provides numerical simulations for benzene and ferrocene.
Section~\ref{sec:demo} presents a hardware demonstration of the proposed method on IBM's quantum hardware.
Conclusions are drawn in Sec.~\ref{sec:conclusion}.

\section{theory}
\label{sec:theory}
\subsection{Finite group and projection operator}
First we consider a general finite symmetry group $G$.
The projection of a general quantum state $|\Psi\rangle$ onto 
one of the irreducible representations $\Gamma$
is~\cite{bishop,10.1063/1.5110682,PhysRevA.111.052433}
\begin{equation}
    P_{\Gamma} |\Psi\rangle = \frac{d_{\Gamma}}{|G|} \sum_{C} \chi^{*}_{\Gamma}(C) \sum_{g \in C}
    \hat{g} |\Psi\rangle,
    \label{eq:p_gamma}
\end{equation}
where $d_{\Gamma}$ is the dimension of $\Gamma$,
$|G|$ is the order of $G$,
$C$ is a conjugacy class with $\chi_{\Gamma}(C)$
its character in $\Gamma$,
and $\hat{g}$ is a symmetry group operation acting on $|\Psi\rangle$.
The state $|\Psi\rangle$ is projected onto $\Gamma$ as
\begin{equation}
 P_{\Gamma}|\Psi\rangle = a_{\Gamma} |\Psi_{\Gamma}\rangle,
\end{equation}
where $|\Psi_{\Gamma}\rangle$ is
a superposition of the basis functions of irreducible representation $\Gamma$,
and $a_{\Gamma}$ is its coefficient. These states are orthonormalized as
$\langle \Psi_{\Gamma}|\Psi_{\Gamma'}\rangle=\delta_{\Gamma \Gamma'}$,
and therefore $\sum_{\Gamma} |a_{\Gamma}|^{2} = 1$.
Note that $P_{\Gamma}$ in Eq.~(\ref{eq:p_gamma})
is related to a more general operator~\cite{bishop}
\begin{equation}
  P_{\Gamma,jj'} = \frac{d_{\Gamma}}{|G|} \sum_{g} \bigl[
    D_{\Gamma}(g)\bigr]^{*}_{jj'} \hat{g} \quad j,j'=1,2,\dots,d_{\Gamma},
  \label{eq:p_gamma_gen}
\end{equation}
where $D_{\Gamma}(g)$ is the irreducible representation matrix for $g$
in $\Gamma$, which is related to $\chi_{\Gamma}(C)$
as $\chi_{\Gamma}(C) = \textrm{Tr} D_{\Gamma}(g \in C)$.
It can be shown that
the state $|\Psi_{\Gamma,jj'}\rangle = P_{\Gamma,jj'}|\Psi\rangle$
behaves as a
$j$-th symmetry-adapted basis function of $\Gamma$~\cite{bishop}.

The central quantity in this paper is the weight of $\Gamma$, defined as $w_{\Gamma} = |a_{\Gamma}|^{2}$. 
This quantity characterizes symmetric properties of the input state $|\Psi\rangle$, and may be viewed as
the power spectrum of the generalized Fourier transform associated with finite group $G$~\cite{PhysRevA.111.052433,10.1145/1198513.1198525}.
Using Eq.~(\ref{eq:p_gamma}), $w_{\Gamma}$ is calculated as
\begin{eqnarray}
  w_{\Gamma} &=& \langle \Psi |P_{\Gamma}|\Psi\rangle \nonumber\\
  &=& \frac{d_{\Gamma}}{|G|} \sum_{C} \chi_{\Gamma}^{*}(C)
  \sum_{g \in C} \langle \Psi | \hat{g}|\Psi\rangle.
  \label{eq:w_gamma}
\end{eqnarray}
In Appendix, we show that in the above expression $\sum_{\Gamma} w_{\Gamma} = 1$ is guaranteed as long as $\langle \Psi |\Psi \rangle = 1$.

\subsection{Point group symmetry in quantum chemistry}
A point group is a fundamental finite symmetry group of a molecule, and is
used to classify its molecular orbitals and electronic states in quantum chemistry.
By operating a point-group symmetry element $\hat{g}$,
a general set of one-particle orbitals $\{\phi_{\nu \sigma}\}$ transform as
\begin{equation}
    \hat{g} \phi_{\nu \sigma} = \sum_{\mu = 1}^{n} \bigl[ D(g) \bigr]_{\mu\nu} \phi_{\mu \sigma},
    \label{eq:g_psi_general}
\end{equation}
where $D(g)$ is the unitary representation matrix of $g$ with respect to $\{ \phi_{\nu\sigma}\}$,
$\mu$ and $\nu$ are spatial orbital indices, and $\sigma\in \{\uparrow, \downarrow\}$ is a spin index.
We restrict ourselves to the spin-restricted case (i.e., $\phi_{\nu \uparrow} = \phi_{\nu \downarrow}$);
the generalization to the spin-unrestricted case is straightforward.

In practical calculations, the representation matrices 
\begin{equation}
  \bigl[ D(g)\bigr]_{\mu \nu} = \langle \phi_{\mu \sigma} |\hat{g}|\phi_{\nu \sigma}\rangle
\end{equation}
need to be prepared. When the orbitals are expanded by some basis functions $\{\mathcal{B}_{I}\}$ as
$\phi_{\nu \sigma} = \sum_{I} x_{I \nu \sigma} \mathcal{B}_{I}$, the matrix elements of $D(g)$ are calculated as
\begin{equation}
  \bigl[ D(g)\bigr]_{\mu \nu} = \sum_{IJK}
  x^{*}_{I \mu \sigma} \mathcal{S}_{IJ} \Bigl[D^{(B)}(g)\Bigr]_{JK} x_{K \nu \sigma},
\end{equation}
where $\mathcal{S}_{IJ} = \langle \mathcal{B}_{I}|\mathcal{B}_{J}\rangle$
and $D^{(B)}(g)$  are the overlap  and the representation matrices for $\{\mathcal{B}_{I}\}$, respectively.

When $\{\phi_{\nu \sigma}\}$ are chosen to be the solution of a one-particle
Hamiltonian having the point group symmetry of the system,
each degenerate set of orbitals form the basis functions of an irreducible representation
of the point group.
Examples of such mean-field-like Hamiltonians include
the Fock operator in the Hartree-Fock approximation~\cite{doi:https://doi.org/10.1002/9781119019572}
and the Kohn-Sham Hamiltonian in density functional theory~\cite{dreizler2012density}.
 The $\gamma$-th set of degenerate orbitals
$S_{\gamma \sigma}=\{\psi_{\gamma j \sigma}\} (j=1,2,\dots,|S_{\gamma \sigma}|)$ hybridize
exclusively with one another, as
\begin{equation}
  \hat{g} \psi_{\gamma k \sigma} = \sum_{j=1}^{|S_{\gamma \sigma}|}
  \bigl[ \bar{D}_{\gamma} (g) \bigr]_{j k} \psi_{\gamma j \sigma}.
  \label{eq:g_psi_irrep}
\end{equation}
In this case $D(g)$ in Eq.~(\ref{eq:g_psi_general})
can be written in a block-diagonal form, and
the size of $\bar{D}_{\gamma}(g)$ in Eq.~(\ref{eq:g_psi_irrep}) is $O(1)$
for simple monomer molecules considered in this work.

Many-electron wavefunctions are expanded by a tensor product of one-particle orthonormal wavefunctions
\begin{equation}
  \sum_{\sigma_{1}\sigma_{2}\dots}\sum_{\nu_{1}\nu_{2}\dots}
  C^{\sigma_{1}\sigma_{2}\dots}_{\nu_{1} \nu_{2} \dots}
  \phi_{\nu_{1}\sigma_{1}}\phi_{\nu_{2}\sigma_{2}} \dots,
\end{equation}
and the reducible representation of 
$g$ with respect to these tensor product basis states is obtained as
\begin{eqnarray}
  \hat{g} \phi_{\nu_{1}\sigma_{1}}\phi_{\nu_{2}\sigma_{2}} \dots &=& \sum_{\mu_{1} \mu_{2}\dots}
  \bigl[ D(g) \bigr]_{\mu_{1} \nu_{1}}
  \bigl[ D(g) \bigr]_{\mu_{2} \nu_{2}}
  \dots \nonumber\\
  && \qquad
  \phi_{\mu_{1}\sigma_{1}} \phi_{\mu_{2}\sigma_{2}} \dots
  \label{eq:g_manybody}
\end{eqnarray}

\subsection{Point group symmetry operations on a quantum computer}
\label{sec:qc_impl}
\begin{figure*}
\includegraphics{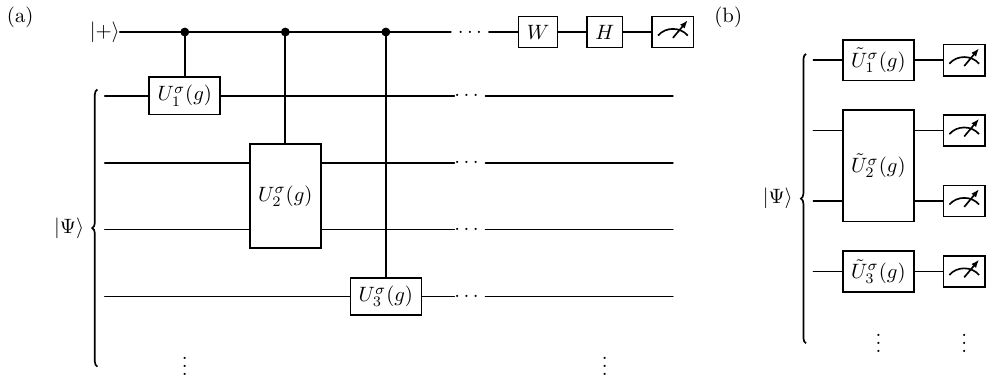}%
\caption{Circuit diagrams for calculating $\langle \Psi |U(g)|\Psi \rangle$ via
the Hadamard test (a) and the proposed ancilla-free method (b). Here $W = I, S^{\dagger}$ for
$\textrm{Re} \langle \Psi |U(g)|\Psi \rangle$ and
$\textrm{Im} \langle \Psi |U(g)|\Psi \rangle$, respectively.
}
\label{fig:expval_ug}
\end{figure*}
As discussed by Yen \textit{et al.}~\cite{10.1063/1.5110682},
a point-group symmetry operation on many-electron states, $\hat{g} |\Psi \rangle = U(g) |\Psi \rangle$,
corresponding to Eq.~(\ref{eq:g_manybody}),
is expressed as an orbital rotation
\begin{eqnarray}
  U(g) &=& \prod_{\sigma} U^{\sigma}(g), \\
  U^{\sigma}(g) &=&
  \exp \Bigl[\sum_{\mu,\nu=1}^{n} \bigl[ \log D(g)\bigr]_{\mu \nu}
  c^{\dagger}_{\mu \sigma} c_{\nu \sigma} \Bigr].
  \label{eq:ug}
\end{eqnarray}
Here $c^{\dagger}_{\mu \sigma}$ and $c_{\nu \sigma}$ are electron creation and annihilation operators, respectively,
and $D(g)$ is defined in Eq.~(\ref{eq:g_psi_general}).
Under the Jordan-Wigner transformation,
this rotation can be implemented using the Givens rotations~\cite{doi:10.1126/science.abb9811}
with $O(n)$ depth for general orbitals (Eq.~(\ref{eq:g_psi_general})).
For symmetry-adapted, mean-field orbitals (Eq.~(\ref{eq:g_psi_irrep})), the depth is reduced to $O(1)$
by decomposing $U^{\sigma}(g)$ as 
\begin{equation}
  U^{\sigma}(g) = \prod_{\gamma} U^{\sigma}_{\gamma}(g),
  \label{eq:ug_ug_gamma}
\end{equation}
where
\begin{equation}
  U^{\sigma}_{\gamma}(g) = 
  \exp \Bigl[\sum_{j,k = 1}^{|S_{\gamma \sigma}|} \bigl[ \log \bar{D}_{\gamma}(g)\bigr]_{j k}
  c^{\dagger}_{\gamma j \sigma} c_{\gamma k \sigma} \Bigr]
  \label{eq:ug_gamma}
\end{equation}
with $\bar{D}_{\gamma}(g)$ defined in Eq.~(\ref{eq:g_psi_irrep}).

Using these operators,
the projection operation (Eq.~(\ref{eq:p_gamma})) is implemented using the linear combination of unitary (LCU)
technique~\cite{Lacroix2023}.
The standard LCU requires
$\lceil \log_{2} |G| \rceil$ ancilla qubits, and
a single-ancilla LCU approach~\cite{Chakraborty2024implementingany} is also proposed for evaluating
expectation values of observables.

\subsection{Weight evaluation on a quantum computer}
Here we consider a quantum-classical hybrid approach to calculate
$w_{\Gamma}$ defined in Eq.~(\ref{eq:w_gamma}),
which evaluates $\langle \Psi | \hat{g} |\Psi \rangle = \langle \Psi | U(g) |\Psi \rangle$
for each $g$ on a quantum computer
and post-processes the results to construct $w_{\Gamma}$.
As discussed in Appendix,
$\sum_{\Gamma} w_{\Gamma} = 1$ is
ensured even in noisy simulations,
provided that $\langle \Psi |U(E)|\Psi\rangle = \langle \Psi|\Psi\rangle = 1$ is given exactly.
It should be noted that $w_{\Gamma}$ can be evaluated in a fully quantum way
by, for example,
measuring ancilla qubits in the generalized symmetry-adapted transform (GSA) proposed in Ref.~\cite{PhysRevA.111.052433},
but these coherent approaches require a deeper circuit with multiply controlled operations.

One way to evaluate $\langle \Psi | U(g) | \Psi \rangle$
is the Hadamard test using one ancilla qubit, as shown
in Fig.~\ref{fig:expval_ug}(a). This approach, however, requires
$O(n)$ long-range controlled $U^{\sigma}_{\gamma}(g)$ operations. Therefore,
it is not suitable for quantum devices with limited qubit connectivity.

To circumvent this difficulty,
we propose an ancilla-free method
shown in Fig.~\ref{fig:expval_ug}(b).
This method is based on the diagonalization
of each $\bar{D}_{\gamma}(g)$
\begin{equation}
  \bar{D}_{\gamma}(g) = V^{g}_{\gamma} \textrm{diag}\Bigl[
    e^{i \varphi^{g}_{\gamma 1}},e^{i \varphi^{g}_{\gamma 2}}, \dots
  \Bigr]
  V^{g \dagger}_{\gamma},
\end{equation}
and rewrites $U^{\sigma}_{\gamma}(g)$ as
\begin{equation}
  U^{\sigma}_{\gamma}(g) = \bigl(\tilde{U}^{\sigma}_{\gamma}(g)\bigr)^{\dagger}
  \Bigl[
  \prod_{k}
   e^{i \varphi^{g}_{\gamma k} \tilde{c}_{\gamma k \sigma}^{\dagger} \tilde{c}_{\gamma k \sigma}}
   \Bigr]
  \tilde{U}^{\sigma}_{\gamma}(g).
  \label{eq:ugnu}
\end{equation}
Here $\tilde{U}^{\sigma}_{\gamma}(g)$ is an orbital rotation
\begin{equation}
  \tilde{U}^{\sigma}_{\gamma}(g) =
  \exp \Bigl[ \sum_{jk} \bigl[\log V^{g \dagger}_{\gamma} \bigr]_{jk}
  c_{\gamma j \sigma}^{\dagger} c_{\gamma k \sigma}\Bigr],
  \label{eq:tilde_ugnu}
\end{equation}
and
\begin{equation}
  \tilde{c}^{\dagger}_{\gamma k \sigma} =
  \sum_{j} \bigl[V^{g}_{\gamma}\bigr]_{j k} c^{\dagger}_{\gamma j \sigma}.
\end{equation}
As $\prod_{k} e^{i \varphi^{g}_{\gamma k} \tilde{c}^{\dagger}_{\gamma k \sigma} \tilde{c}_{\gamma k \sigma}}$
in Eq.~(\ref{eq:ugnu}) is diagonal
in the computational basis,
$\langle \Psi |U(g)|\Psi \rangle$ can be obtained by
measuring
$|\tilde{\Psi}(g)\rangle = \prod_{\sigma} \prod_{\gamma} \tilde{U}^{\sigma}_{\gamma}(g) |\Psi\rangle$
and combining the results with appropriate phase factors.
We note that the qubit tapering technique proposed in Ref.~\cite{Setia2020} employs a similar methodology for abelian groups.
Our approach can also be used for non-abelian groups and for general orbitals (Eq.~(\ref{eq:g_psi_general})) with 
$O(n)$ depth overhead for the orbital rotation.

Next we discuss the sample complexity of the method. Noting
\begin{equation}
\textrm{Var}
\Bigl[
  \prod_{\sigma}\prod_{\gamma}\prod_{k} e^{i \varphi_{\gamma k}^{g} \tilde{c}^{\dagger}_{\gamma k \sigma} \tilde{c}_{\gamma k \sigma}}
\Bigr]
  \leq 1, 
\end{equation}
the number of total measurements $M$
for a given total variance $\epsilon^{2}$
is bounded as~\cite{Rubin_2018}
\begin{equation}
  M \leq \frac{1}{\epsilon^{2}}
  \Bigl(
    \sum_{C} \frac{r_{C}}{|G|} |\chi_{\Gamma}(C)|
  \Bigr)^{2},
  \label{eq:m}
\end{equation}
where $r_{C}$ is defined in Eq.~(\ref{eq:r_c}).
Using the orthonormality of $\chi_{\Gamma}(C)$
(Eq.~(\ref{eq:chiorth1}))
and the Cauchy-Schwarz inequality,
Eq.~(\ref{eq:m}) becomes
\begin{equation}
M \leq \frac{1}{\epsilon^{2}}
\Bigl(\sum_{C}r_{C} \cdot \frac{|\chi_{\Gamma}(C)|^{2}}{|G|^{2}}\Bigr)
\cdot
  \Bigl(\sum_{C} r_{C} \cdot 1^{2} \Bigr)
= \frac{1}{\epsilon^{2}},
\end{equation}
which indicates that the number of measurements required is 
independent of $|G|$ and $\Gamma$.

For abelian groups, all irreducible representations
are one-dimensional. In this case $\tilde{U}_{\gamma}(g) = I$ for all $g \in G$, and therefore all
$U(g)$ can be measured simultaneously.
Furthermore, when $\bar{D}_{\gamma}(g) = \pm 1$ for all $g$,
$\langle \Psi |U(g)|\Psi \rangle$ can be obtained by
calculating the expectation value of a single Pauli operator
$P(g) = \prod_{\sigma} \prod_{\gamma} P_{\gamma}(g)$,
where
\begin{equation}
  P_{\gamma}(g) = \left\{
    \begin{array}{ll}
      I & \quad \bar{D}_{\gamma}(g) = 1 \\
      Z & \quad \bar{D}_{\gamma}(g) = -1 .
    \end{array}
    \right.
\end{equation}

\section{Numerical simulations on a classical computer}
\label{sec:num}
\begin{figure*}
\includegraphics{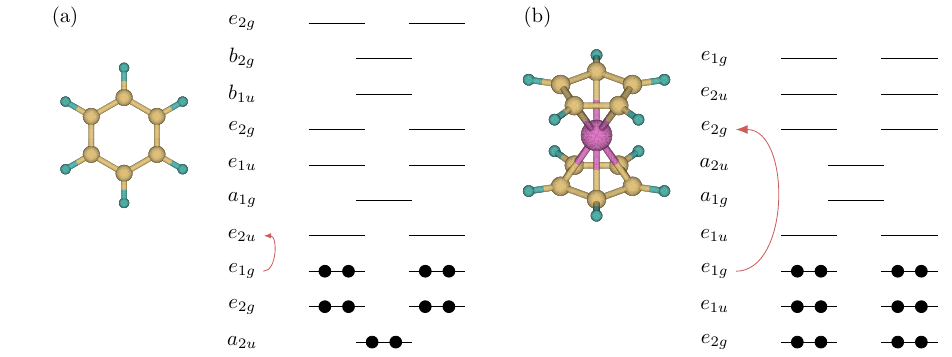}%
\caption{Structures and Hartree-Fock one-particle energy diagrams of benzene (a) and staggered ferrocene (b).
Filled circles denote occupied electrons in the Hartree-Fock ground-state configuration.
Arrows indicate single excitations considered in this work.
}
\label{fig:levels}
\end{figure*}
\begin{table*}
  \centering
  \setlength{\tabcolsep}{5pt}
  \begin{tabular}{ccccccccccccc}
    \hline
    $D_{6h}$ & $E$ & $2 C_{6}$ & $2C_{3}$ & $C_{2}''$ & $3C_{2}$ & $3C_{2}'$ & $\sigma_{h}$
    & $3 \sigma_{v}$ & $3 \sigma_{d}$ & $2S_{6}$ & $2 S_{3}$ & $i$ \\ \hline
    $A_{1g}$ & +1 & +1 & +1 & +1 & +1 & +1 &
               +1 & +1 & +1 & +1 & +1 & +1 \\
    $A_{1u}$ & +1 & +1 & +1 & +1 & +1 & +1 &
               -1 & -1 & -1 & -1 & -1 & -1 \\
    $A_{2g}$ & +1 & +1 & +1 & +1 & -1 & -1 &
               +1 & -1 & -1 & +1 & +1 & +1 \\
    $A_{2u}$ & +1 & +1 & +1 & +1 & -1 & -1 &
               -1 & +1 & +1 & -1 & -1 & -1 \\
    $B_{1g}$ & +1 & -1 & +1 & -1 & +1 & -1 &
               -1 & -1 & +1 & +1 & -1 & +1 \\
    $B_{1u}$ & +1 & -1 & +1 & -1 & +1 & -1 &
               +1 & +1 & -1 & -1 & +1 & -1 \\
    $B_{2g}$ & +1 & -1 & +1 & -1 & -1 & +1 &
               -1 & +1 & -1 & +1 & -1 & +1 \\
    $B_{2u}$ & +1 & -1 & +1 & -1 & -1 & +1 &
               +1 & -1 & +1 & -1 & +1 & -1 \\
    $E_{1g}$ & +2 & +1 & -1 & -2 &  0 &  0 &
               -2 &  0 &  0 & -1 & +1 & +2 \\
    $E_{1u}$ & +2 & +1 & -1 & -2 &  0 &  0 &
               +2 &  0 &  0 & +1 & -1 & -2 \\
    $E_{2g}$ & +2 & -1 & -1 & +2 &  0 &  0 &
               +2 &  0 &  0 & -1 & -1 & +2 \\
    $E_{2u}$ & +2 & -1 & -1 & +2 &  0 &  0 &
               -2 &  0 &  0 & +1 & +1 & -2 \\
             \hline
  \end{tabular}
  \caption{Character table of $D_{6h}$.}
  \label{tbl:d6h}
\end{table*}
\begin{table}
  \centering
  \begin{tabular}{ccccccccc}
    \hline
    $D_{5d}$ & $E$ & $2C_{5}$ & $2C_{5}^{2}$ & 5$C_{2}'$ & $i$
    & $2S_{10}^{3}$ & $2S_{10}$ & $5\sigma_{d}$ \\ \hline
    $A_{1g}$ & +1 & +1 & +1 & +1 &
               +1 & +1 & +1 & +1 \\
    $A_{1u}$ & +1 & +1 & +1 & +1 &
               -1 & -1 & -1 & -1 \\
    $A_{2g}$ & +1 & +1 & +1 & -1 &
               +1 & -1 & +1 & -1 \\
    $A_{2u}$ & +1 & +1 & +1 & -1 &
               -1 & -1 & -1 & +1 \\
    $E_{1g}$ & +2 & $+x_{+}$ & $+x_{-}$ &  0 &
               +2 & $+x_{+}$ & $+x_{-}$ &  0 \\
    $E_{1u}$ & +2 & $+x_{+}$ & $+x_{-}$ &  0 &
               -2 & $-x_{+}$ & $-x_{-}$ &  0 \\
    $E_{2g}$ & +2 & $+x_{-}$ & $+x_{+}$ &  0 &
               +2 & $+x_{-}$ & $+x_{+}$ &  0 \\
    $E_{2u}$ & +2 & $+x_{-}$ & $+x_{+}$ &  0 &
               -2 & $-x_{-}$ & $-x_{+}$ & +1 \\
             \hline
  \end{tabular}
  \caption{Character table of $D_{5d}$. Here $x_{\pm}=\frac{-1 \pm \sqrt{5}}{2}$.}
  \label{tbl:d5d}
\end{table}

For a better understanding of point-group
symmetry properties of many-electron wavefunctions, we present
classical simulations on two prototype molecules,
benzene (C$_{6}$H$_{6}$) and
staggered ferrocene (FeC$_{10}$H$_{10}$)~\cite{Sohn-1971, MOHAMMADI201251}, which belong to
$D_{6h}$ and $D_{5d}$ point groups, respectively.
The character tables of
these two point groups are shown in Tables~\ref{tbl:d6h} and
\ref{tbl:d5d}.

The calculations are conducted using ffsim~\cite{sung2026ffsimfastersimulationfermionic,ffsim}, and 
the Hamiltonian and the representation matrix elements are
calculated using pyscf~\cite{10.1063/5.0006074} with 6-31G basis. 
The bond lengths of benzene used are 1.39 \AA{ }  and 1.09 \AA{ } for C-C and C-H bonds,
respectively.
The structure of the staggered ferrocene is taken from Ref.~\cite{MOHAMMADI201251}.
The active spaces of benzene and ferrocene are chosen
as (10e,16o) and (12e,16o), respectively. Here,
($m$e,$n$o) denotes a space spanned by $m$ electrons in $n$ spatial orbitals.
The structures and the Hartree-Fock one-particle energy diagrams of the two molecules are
shown in Fig.~\ref{fig:levels}, together with the irreducible representation labels (in lowercase)
of the orbitals.
\subsection{Single Slater determinant case}

\begin{figure}
\includegraphics{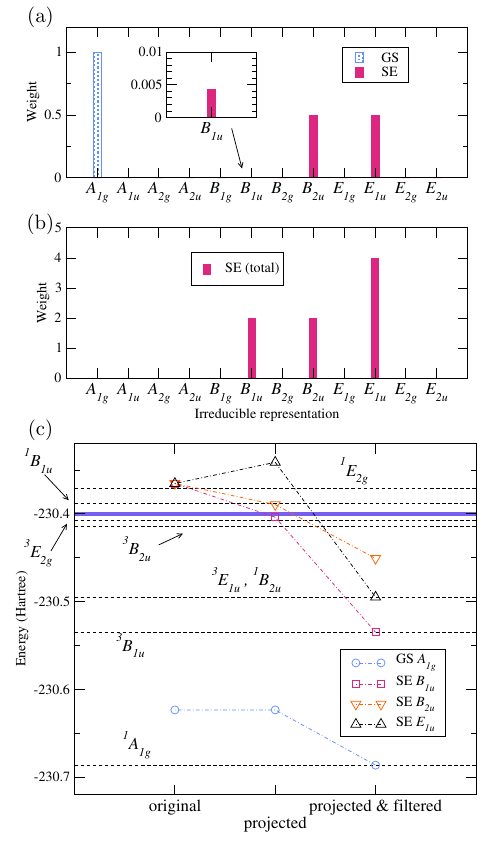}%
\caption{(a) Weights $w_{\Gamma}$ for the Hartree-Fock ground-state (GS)
and one single-excited state (SE) configurations of benzene. The inset shows a magnified view.
(b) Sum of weights over the eight SE configurations (Eq.~(\ref{eq:w_se_total})).
(c) Energy expectation values of (i) original Slater determinant states,
(ii) symmetry-projected states,
and (iii) symmetry-projected and filtered states. The thick horizontal line shows the energy cutoff of the filter,
and the dashed lines show the exact eigenenergies.
The superscripts of the eigenstate labels denote total spin $2S + 1$.
Note that $^3 E_{1u}$ and $^1 B_{2u}$ eigenenergies
are very close and indistinguishable on this scale.}
\label{fig:benzene_ssd}
\end{figure}

\begin{figure}
\includegraphics{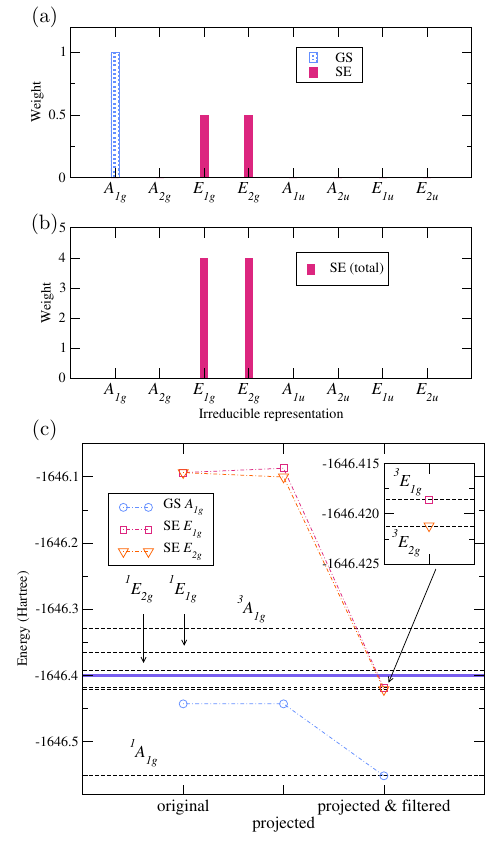}%
\caption{Same as Fig.~\ref{fig:benzene_ssd} but for ferrocene.}
\label{fig:ferrocene_ssd}
\end{figure}

We start with the case where
the wavefunction is expressed as a single Slater determinant
\begin{equation}
    |\Psi\rangle =  \prod_{\sigma  \in \{\uparrow, \downarrow\}} \prod_{\mu_{\sigma}=1}^{n_{\sigma}}
    c^{\dagger}_{p_{\mu_{\sigma}} \sigma} |0\rangle,
\end{equation}
where $n_{\sigma}$ is the number of electrons in a spin state $\sigma$, and
$\{p_{\mu_{\sigma}}\}$ denote occupied orbitals.
In this case $\langle\Psi| U(g) |\Psi\rangle$ is easily calculated as
\begin{equation}
    \langle\Psi| U(g) |\Psi\rangle = \prod_{\sigma \in \{\uparrow, \downarrow\}} \textrm{det} \mathcal{D}_{\sigma}(g),
    \label{eq:ug_sd}
\end{equation}
where $\mathcal{D}_{\sigma}(g)$ is an $n_{\sigma} \times n_{\sigma}$
matrix whose matrix elements are defined as
\begin{equation}
    \bigl[
    \mathcal{D}_{\sigma}(g)\bigr]_{\mu_{\sigma} \nu_{\sigma}} = \bigl[D(g)\bigr]_{p_{\mu_{\sigma}} p_{\nu_{\sigma}}}.
\end{equation}

We calculate the weights $w_{\Gamma}$ for the Hartree-Fock ground states and
single excited (SE) states of the two molecules.
This calculation is inspired by the work in Ref.~\cite{doi:10.1021/ed048p92},
where the reduction of point-group representations formed from
many-electron wavefunctions in benzene and xenon tetrafluoride (XeF$_4$) is carried out.
The Hartree-Fock ground-state configurations are
$(a_{2u})^{2}(e_{2g})^{4}(e_{1g})^{4}$ and
$(e_{2g})^{4}(e_{1u})^{4}(e_{1g})^{4}$
for benzene and ferrocene, respectively (see Fig.~\ref{fig:levels}).
The SE configurations considered are 
$(a_{2u})^{2}(e_{2g})^{4}(e_{1g})^{3}(e_{2u})^{1}$ and
$(e_{2g})^{4}(e_{1u})^{4}(e_{1g})^{3} (e_{2g})^{1}$, respectively,
as indicated by the arrows in Fig.~\ref{fig:levels}.
For each molecule,
there are eight independent SE configurations with $S_{z}=0$, where $S_{z}$ is the $z$-component of the total spin.

The calculated weights for benzene and ferrocene are plotted
in Fig.~\ref{fig:benzene_ssd}(a) and Fig.~\ref{fig:ferrocene_ssd}(a), respectively.
For the SE states, the results of one configuration (out of eight) are shown for each
molecule.
The Hartree-Fock ground states of both molecules belong to the perfectly symmetric $A_{1g}$
irreducible representation,
as expected for any closed shell configuration~\cite{doi:10.1021/ed048p92}.
On the other hand, the SE states in both molecules
consist of multiple irreducible representations.
For benzene, shown in Fig.~\ref{fig:benzene_ssd}(a),
the chosen SE state
is decomposed as $\approx 0.005 B_{1u} + 0.495 B_{2u} + 0.5 E_{1u}$.
Note that in this partially-filled configuration
these values are dependent on the details of the diagonalization routine
used in the Hartree-Fock calculation, as they are not invariant with respect
to the unitary transformation of the degenerate orbitals
\begin{equation}
\psi_{\gamma k \sigma} \to \psi'_{\gamma k \sigma} = \sum_{j} V^{\gamma}_{jk}
\psi_{\gamma j \sigma}
\label{eq:u_trans}
\end{equation}
with $V^{\gamma}_{j k}$ an arbitrary unitary matrix.

The reduction of the eight-dimensional representation for the SE states, similar to
Ref.~\cite{doi:10.1021/ed048p92}, can be done by
summing the weights for all the eight SE configurations, namely, by calculating
\begin{eqnarray}
w^{\textrm{SE-total}}_{\Gamma} &=& \sum_{i \in \textrm{SE}} w^{(i)}_{\Gamma}\nonumber\\
 &=& \frac{d_{\Gamma}}{|G|} \sum_{C} \chi^{*}_{\Gamma}(C) \sum_{g\in C} \nonumber\\
 && \qquad \sum_{i \in \textrm{SE}} \langle \Psi^{(i)}|U(g)|\Psi^{(i)}\rangle .
 \label{eq:w_se_total}
\end{eqnarray}
Here $|\Psi^{(i)}\rangle$ is the $i$-th SE configuration.
Note that 
$\sum_{i \in \textrm{SE}} \langle \Psi^{(i)}|U(g)|\Psi^{(i)}\rangle$ corresponds
to the character of the (reducible) representation for the SE states,
and Eq.~(\ref{eq:w_se_total}) coincides with the well-known formula for counting the
number of occurrences of $\Gamma$ in a given representation~\cite{doi:10.1021/ed048p92},
up to a prefactor of $d_{\Gamma}$.
As can be seen in Fig.~\ref{fig:benzene_ssd}(b),
this eight-dimensional space in benzene is decomposed as $2 B_{1u} + 2 B_{2u} + 4 E_{1u}$,
and this result is invariant with respect to the orbital transformation (Eq.~(\ref{eq:u_trans})).
Similarly, one SE configuration of ferrocene
is decomposed as $0.5 E_{1g} + 0.5 E_{2g}$, and the total
eight-dimensional SE space
is decomposed as $4 E_{1g} + 4 E_{2g}$, as shown in Fig.~\ref{fig:ferrocene_ssd}(b).

To investigate the effect of the symmetry projection (Eq.~(\ref{eq:p_gamma})),
in Fig.~\ref{fig:benzene_ssd}(c) and Fig.~\ref{fig:ferrocene_ssd}(c)
the energy expectation values of the following three wavefunction sets are calculated for
the two molecules:

\begin{enumerate}
  \item the original Hartree-Fock ground state and one SE state.
  \item states obtained by applying the symmetry projection $P_{\Gamma}$ to 1.
  \item states obtained by applying an energy-based filter to 2.
\end{enumerate}
We apply a step-function like filter with cutoff energy as
$\Theta (H - E_{\textrm{cutoff}}) |\Psi\rangle$,
where
\begin{equation}
  \Theta (E)  = \left\{
\begin{array}{ll}
  1 & E < 0 \\
  0 & E > 0.
\end{array}
  \right.
\end{equation}
This filter can be implemented, for example,
with the quantum eigenvalue transformation of unitary matrices (QETU)~\cite{PRXQuantum.3.040305}.
In Fig.~\ref{fig:benzene_ssd}(c) and Fig.~\ref{fig:ferrocene_ssd}(c),
the cutoff energy and the exact eigenstates are shown by thick and dashed lines, respectively.

In both cases, applying the symmetry projection to the Hartree-Fock ground state
does not decrease the energy,
as the state  already belongs to the correct irreducible representation ($A_{1g}$) of the ground state.
For the SE states, their energy expectation values after the symmetry projection
are also higher than the lowest exact eigenenergies with the same irreducible representations.
By applying the filter, the exact eigenenergies are obtained for all the projected states
except the SE state for benzene projected onto $B_{2u}$ (Fig.~\ref{fig:benzene_ssd}(c)).
The reason for this discrepancy is that there are two eigenstates below the cutoff belonging to $B_{2u}$
but with different spin ($^{1}B_{2u}$ and $^{3}B_{2u}$); this example shows the limitation
of the current procedure.

This simple
numerical experiment confirms that the symmetry projection (Eq.~(\ref{eq:p_gamma})) is useful
to get exact eigenstates with specific irreducible representations, but
it has to be combined with other algorithms, such as filtering~\cite{PRXQuantum.3.040305},
quantum phase estimation~\cite{kitaev1995quantummeasurementsabelianstabilizer},
or spin projection~\cite{10.1063/1.5110682}.
Our proposed method provides a simple yet powerful tool for analyzing symmetries of a trial wavefunction
before performing the projection.

\subsection{Interacting case: local unitary cluster Jastrow function}
\begin{figure}
\includegraphics{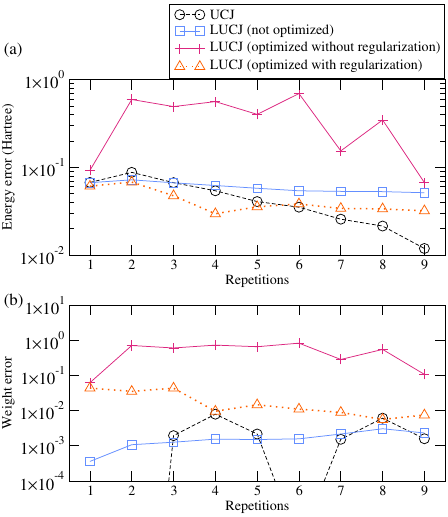}%
\caption{Errors in the ground-state energy (a) and the $A_{1g}$ weight (b) of the UCJ and LUCJ functions in benzene.}
\label{fig:benzene_ucj}
\end{figure}

\begin{figure}
\includegraphics{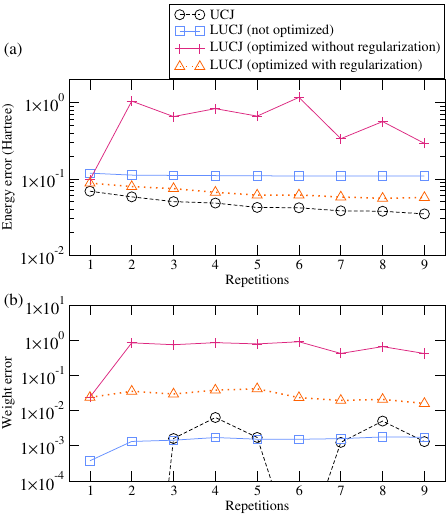}%
\caption{Errors in the ground-state energy (a) and the $A_{1g}$ weight (b) of the UCJ and LUCJ functions in ferrocene.}
\label{fig:ferrocene_ucj}
\end{figure}
As an example of correlated wavefunctions,
we next present a symmetry analysis of
the unitary cluster Jastrow (UCJ) function and
its local variant known as the local unitary cluster Jastrow (LUCJ) function~\cite{D3SC02516K}.
These wavefunctions are proposed as an approximation of 
the unitary coupled-cluster wavefunction,
and are often used as a trial state of
the sample-based quantum diagonalization~\cite{doi:10.1126/sciadv.adu9991}.

The UCJ wavefunction has the form~\cite{D3SC02516K,doi:10.1126/sciadv.adu9991,lin2025improvedparameterinitializationlocal}
\begin{equation}
  |\Psi\rangle = \prod_{r=1}^{R} \mathcal{U}_{r} e^{i \mathcal{J}_{r}} \mathcal{U}_{r}^{\dagger}
  |\Psi_{\textrm{0}}\rangle,
\end{equation}
where $|\Psi_{\textrm{0}}\rangle$ is a reference state
which we take as the Hartree-Fock ground state,
$\mathcal{U}_{r}$ is an orbital rotation, $R$ is the number of repetitions,
and $\mathcal{J}_{r}$ is a density-density interaction operator
\begin{equation}
  \mathcal{J}_{r} = \frac{1}{2} \sum_{\sigma,\sigma'}\sum_{\mu,\nu}
  J^{(r)\sigma \sigma'}_{\mu \nu} n_{\mu \sigma} n_{\nu \sigma'}
  \label{eq:j}
\end{equation}
with $n_{\mu\sigma} = c^{\dagger}_{\mu \sigma} c_{\mu \sigma}$.

The LUCJ wavefunction is obtained by retaining only
selected terms in Eq.~(\ref{eq:j}) that are compatible with
a specific qubit connectivity.
For the heavy-hexagonal connectivity considered in this work,
$J^{(r)\sigma \sigma'}_{\mu \nu}$ is allowed to be nonzero
only for the following orbital sets:
\begin{eqnarray}
  S_{\sigma = \sigma'} &=& \{(p, p+1)| p=1,2,\dots,n - 1\} \\
  S_{\sigma \neq \sigma'} &=& \{(p, p)| p=1, 5, 9,\dots, n\}.
\end{eqnarray}

The parameters in the (L)UCJ wavefunctions are often
initialized using the truncated double-factorized form
of the $t_{2}$ amplitudes obtained from a classical coupled-cluster singles and doubles (CCSD) calculation
~\cite{doi:10.1126/sciadv.adu9991,lin2025improvedparameterinitializationlocal}
\begin{equation}
  t_{oo' uu'} \approx i \sum_{r=1}^{R}\sum_{\mu\nu}
  J^{(r)}_{\mu\nu} U_{u\mu}^{(r)}U_{o\mu}^{(r)*}U_{u'\nu}^{(r)}U_{o'\nu}^{(r)*}.
\end{equation}
Here $o$ and $o'$ ($u$ and $u'$) denote occupied (unoccupied)
orbitals, $J^{(r)}$ is a real symmetric matrix
used in the interaction operator (Eq.~(\ref{eq:j})),
and $U^{(r)}$ is a unitary matrix for an orbital rotation.

In Ref.~\cite{lin2025improvedparameterinitializationlocal}, two methods
are proposed to optimize the parameters in the (L)UCJ functions, and
we focus on their first method. This method optimizes
the sparsified interaction matrices $\bar{J}^{(r)}$ due to connectivity restriction
and the corresponding rotation matrices
$\bar{U}^{(r)}$
by minimizing
\begin{equation}
  \frac{1}{2}\sum_{oo'uu'} |\bar{t}_{oo'uu'} - t_{oo'uu'}|^{2} + \lambda
  \Bigl| \sum_{r} || \bar{J}^{(r)}||_{\textrm{F}}^{2}
  -
  \sum_{r} || J^{(r)}||_{\textrm{F}}^{2}
  \Bigr|.
  \label{eq:min_j}
\end{equation}
Here $\bar{t}_{oo'uu'}$ are compressed $t_{2}$ amplitudes~\cite{lin2025improvedparameterinitializationlocal} constructed
from $\bar{J}^{(r)}$ and $\bar{U}^{(r)}$,
and the subscript $\textrm{F}$ denotes the Frobenius norm.
The second term in Eq.~(\ref{eq:min_j})
is a regularization term used to prevent one term from becoming too large.

Our objective here is to investigate the point-group symmetry properties of the UCJ and LUCJ functions
and see the effect of the optimization procedure described above.
We prepare the (L)UCJ functions for benzene and ferrocene using ffsim~\cite{sung2026ffsimfastersimulationfermionic,ffsim}, and
calculate the energy expectation value
and also
the weight of $A_{1g}$, which is the irreducible representation of the exact ground state.

The calculated results as a function of repetitions $R$
are plotted in Fig.~\ref{fig:benzene_ucj} and Fig.~\ref{fig:ferrocene_ucj}
for benzene and ferrocene, respectively.
For the LUCJ functions, we compare three parameter sets:
(i) unoptimized (initialized with the CCSD $t_{2}$ amplitudes),
(ii) optimized without regularization, and (iii) optimized
with regularization parameter $\lambda = 0.005$~\cite{lin2025improvedparameterinitializationlocal}.
For comparison, the results for the UCJ function without parameter optimization are also plotted.

As also reported in Ref.~\cite{lin2025improvedparameterinitializationlocal}, in both
systems the optimized LUCJ wavefunction
without regularization results in the largest energy error. The error in the weight of $A_{1g}$
for this wavefunction is also significantly large in most cases,
indicating that the optimization through Eq.~(\ref{eq:min_j}) does not preserve the symmetry of the reference wavefunction.
For the other three wavefunctions, the discrepancy is smaller. Perhaps not surprisingly,
the UCJ and unoptimized LUCJ wavefunctions
provide very high ($>99$\%) $A_{1g}$ weights,
as the $t_{2}$ amplitudes in CCSD reflect the symmetries of the orbitals.
It can also be seen that
the optimized LUCJ wavefunction with regularization yields a smaller $A_{1g}$ weight than
the unoptimized LUCJ, but it improves the energy expectation value.
This indicates that having a higher weight of the correct irreducible representation
does not necessarily guarantee a higher-quality approximation.

In Ref.~\cite{lin2025improvedparameterinitializationlocal}, it is also reported that
when used as a trial state for the sample-based energy estimation with
quantum-selected configuration interaction
(QSCI)~\cite{kanno2023quantumselectedconfigurationinteractionclassical},
the (L)UCJ function optimized without regularization yields a lower energy than
those optimized with regularization.
Our symmetry analysis suggests that
using symmetry-broken wavefunctions could enhance the efficiency of sample-based approaches.

\section{Demonstration on real quantum hardware}
\label{sec:demo}
\begin{table}
  \centering
  \begin{tabular}{lr}
    \hline
    Date & April 29, 2026 \\
    Number of qubits & 156 \\
    Processor type & Heron r2 \\
    Basis gates & \texttt{cz,id,rx,rz,rzz,sx,x} \\
    Median readout error & 5.49$\times 10^{-3}$ \\
    Median \texttt{cz} error & 1.677$\times 10^{-3}$\\
    Median \texttt{sx} error & 1.973$\times 10^{-4}$ \\
    Median $T_{1}$ ($\mu$s) & 307.34 \\
    Median $T_{2}$ ($\mu$s) & 157.38 \\
    \hline
  \end{tabular}
  \caption{Hardware specifications of \texttt{ibm\_kawasaki}.}
  \label{tbl:kawasaki}
\end{table}
\begin{table}
  \centering
  \begin{tabular}{ccccccccc}
    \hline
    $D_{2h}$ & $E$ & $C_{2}$ & $C_{2}'$ & $C_{2}''$ & $i$ & $\sigma_{h}$
    & $\sigma_{v}$ & $\sigma_{d}$ \\ \hline
    $A_{g}$ & +1 & +1 & +1 & +1 &
              +1 & +1 & +1 & +1 \\
    $A_{u}$ & +1 & +1 & +1 & +1 &
               -1 & -1 & -1 & -1 \\
    $B_{1g}$ & +1 & +1 & -1 & -1 &
               +1 & +1 & -1 & -1 \\
    $B_{1u}$ & +1 & +1 & -1 & -1 &
               -1 & -1 & +1 & +1 \\
    $B_{2g}$ & +1 & -1 & -1 & +1 &
               +1 & -1 & +1 & -1 \\
    $B_{2u}$ & +1 & -1 & -1 & +1 &
               -1 & +1 & -1 & +1 \\
    $B_{3g}$ & +1 & -1 & +1 & -1 &
               +1 & -1 & -1 & +1 \\
    $B_{3u}$ & +1 & -1 & +1 & -1 &
               -1 & +1 & +1 & -1 \\
             \hline
  \end{tabular}
  \caption{Character table of $D_{2h}$.}
  \label{tbl:d2h}
\end{table}

\begin{table*}
  \centering
  \setlength{\tabcolsep}{5pt}
  \begin{tabular}{ccccccccccccc}
    \hline
    $D_{6h}$ &
    $A_{1g}$ & $A_{1u}$ & $A_{2g}$ & $A_{2u}$ &
    $B_{1g}$ & $B_{1u}$ & $B_{2g}$ & $B_{2u}$ &
    $E_{1g}$ & $E_{1u}$ & $E_{2g}$ & $E_{2u}$ \\ \hline
    $D_{2h}$ &
    $A_{g}$  & $A_{u}$  & $B_{1g}$ & $B_{1u}$ &
    $B_{2g}$ & $B_{2u}$ & $B_{3g}$ & $B_{3u}$ &
    $B_{2g}+B_{3g}$ & $B_{2u}+B_{3u}$ & $A_{g}+B_{1g}$ & $A_{u}+B_{1u}$ \\
    \hline
  \end{tabular}
  \caption{Correspondence between $D_{6h}$ and $D_{2h}$ irreducible representations.}
  \label{tbl:d6h_d2h}
\end{table*}

In this section, we demonstrate the evaluation of
the weights $w_{\Gamma}$ in Eq.~(\ref{eq:w_gamma}) on IBM's superconducting quantum device \texttt{ibm\_kawasaki}
for the ground state
and the first excited state of benzene.
The specifications of the device are summarized in Table~\ref{tbl:kawasaki}.

We consider three active spaces, (10e,8o), (10e,12o), and (10e,16o)
with the Jordan-Wigner encoding, corresponding to 16, 24, and 32 qubits, respectively.
For simplicity, we assume
abelian $D_{2h}$ point group (Table~\ref{tbl:d2h}) instead of $D_{6h}$.
In this reduced point group, the irreducible representations of $D_{6h}$ transform according to Table~\ref{tbl:d6h_d2h}.
We calculate the Hartree-Fock orbitals of benzene, used as the encoding basis,
with imposing $D_{2h}$ symmetry.

We employ the Pauli-based evaluation method explained in Sec.~\ref{sec:qc_impl};
each $\langle \Psi |U(g)|\Psi\rangle$
is evaluated as the expectation value of a single Pauli operator $P(g)$.
In the case of $D_{2h}$,
there are eight mutually commuting Pauli operators, including one identity operation.

\subsection{Compression of DMRG wavefunctions}
\begin{figure*}[ht]
\includegraphics{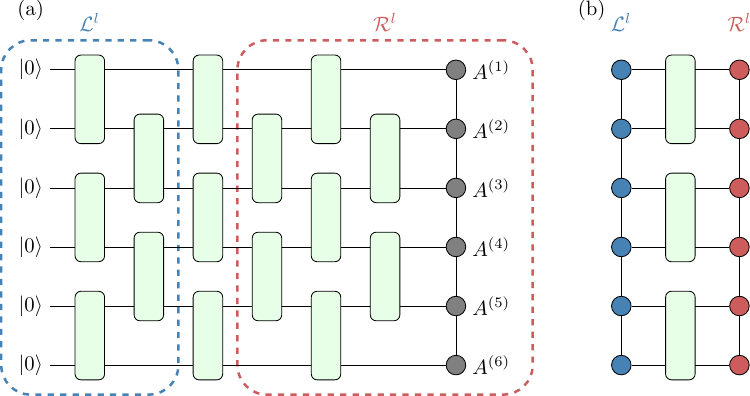}%
\caption{(a) Schematic of our optimization scheme for $L=6$ and $2n=6$.
Rounded rectangles and circles denote two-qubit unitaries $\{U_{lb}\}$
and MPS matrices $\{A^{(p)}\}$ (Eq.~(\ref{eq:mps})), respectively.
(b) To optimize $U_{lb}$ in the $l$-th
layer, two MPSs $\mathcal{L}^{l}$ and $\mathcal{R}^{l}$ are constructed
using $U_{l'\ne l, b}$ inside the two dashed boxes in (a).}
\label{fig:bw_mps}
\end{figure*}
Similarly to Ref.~\cite{doi:10.1073/pnas.2425026122},
we prepare trial wavefunctions from classical DMRG calculations
and convert them into a brick-wall ansatz for approximate circuit encoding.
After the Jordan-Wigner transformation,
the DMRG wavefunction of $2n$ spin orbitals
is written in an MPS form
\begin{eqnarray}
  |\Psi_{\textrm{DMRG}}\rangle &=& \sum_{\sigma_{1},\sigma_{2},\dots,\sigma_{2n}}
  \sum_{m_{2},m_{3},\dots,m_{2n}} \nonumber\\
  &&
  A^{(1)}_{m_{1} \sigma_{1} m_{2}}
  A^{(2)}_{m_{2} \sigma_{2} m_{3}}
  \dots A^{(2n)}_{m_{2n} \sigma_{2n} m_{2n+1}}
  \nonumber\\
  &&\times|\sigma_{1}\sigma_{2}\dots\sigma_{2n}\rangle,
  \label{eq:mps}
\end{eqnarray}
with $m_{1}=m_{2n+1}=1$.
Here $\sigma_{p}$ and $m_{p}$ $(p=1,2,\dots,2n)$
denote physical spin and virtual bond indices, respectively,
and $A^{(p)}_{m_{p}\sigma_{p}m_{p+1}}$ in this work
is right-normalized as
\begin{equation}
\sum_{\sigma_{p} m_{p+1}}
A^{(p) *}_{m_{p}\sigma_{p} m_{p + 1}}
A^{(p)}_{m'_{p}\sigma_{p} m_{p + 1}}
= \delta_{m_{p} m'_{p}}.
\end{equation}
This wavefunction is encoded in a quantum circuit
using a brick-wall ansatz with $L$ layers
\begin{equation}
  |\tilde{\Psi}\rangle = \prod_{l=1}^{L} \prod_{b=1}^{B_{l}} U_{l b} | 0 \rangle^{\otimes 2 n},
\end{equation}
where $U_{l b}$ is a two-qubit unitary, and $B_{l}$ is
the number of $U_{l b}$ in layer $l$
\begin{equation}
  B_{l} = \left\{
    \begin{array}{ll}
      n & \textrm{odd} \,\, l \\
      n-1 & \textrm{even} \,\, l .
    \end{array}
  \right.
\end{equation}

We optimize $U_{lb}$ iteratively using a scheme
similar to Ref.~\cite{doi:10.1073/pnas.2425026122};
we introduce the cost function
\begin{eqnarray}
  \mathcal{C} &=& \bigl|\bigl| \,|\Psi_{\textrm{DMRG}}\rangle
  - |\tilde{\Psi}\rangle \bigr|\bigr|^{2} \nonumber\\
  &=& 2 - 2 \textrm{Re} \langle \Psi_{\textrm{DMRG}}|\tilde{\Psi}\rangle,
  \label{eq:cost_bw_mps}
\end{eqnarray}
and perform a layer-by-layer optimization.
As schematically shown
in Fig.~\ref{fig:bw_mps}, to optimize $U_{lb}$ in layer $l$,
two auxiliary MPSs $\mathcal{L}^{l}$ and 
$\mathcal{R}^{l}$ are prepared.
The ``left'' MPS $\mathcal{L}^{l}$ is calculated
by operating $U_{l' b}$ on $| 0 \rangle^{\otimes 2n}$ sequentially for $l'=1,2,\dots,l-1$,
and converting the resulting state
into an MPS form in each step
via the singular value decomposition (SVD)~\cite{SCHOLLWOCK201196,PhysRevX.10.041038}.
Similarly, the ``right'' MPS $\mathcal{R}^{l}$ is calculated
by applying $U^{\dagger}_{l' b}$ in reverse order, for $l'=L, L-1, \dots, l+1$ to $|\Psi_{\textrm{DMRG}}\rangle$.

The optimal $U_{l b}$ in layer $l$ is
obtained via the SVD~\cite{PhysRevResearch.6.043008,doi:10.1073/pnas.2425026122};
to be specific, Eq.~(\ref{eq:cost_bw_mps}) is rewritten as
\begin{equation}
  \mathcal{C} = 2 - 2 \textrm{Re} \textrm{Tr}\Bigl[ \mathcal{E}^{\dagger}_{lb} U_{lb}\Bigr],
\end{equation}
where $\mathcal{E}_{lb}$ is the environment matrix obtained by contracting $\mathcal{L}^{l}$
$\mathcal{R}^{l}$, and $\{U_{l b' \neq b}\}$. This matrix is decomposed as
\begin{equation}
  \mathcal{E}_{lb} = \mathcal{U} \Sigma \mathcal{V}^{\dagger},
\end{equation}
where $\mathcal{U}, \mathcal{V}$ are unitary matrices, and $\Sigma$ is a diagonal matrix
with singular values on the diagonal.
The optimal $U_{lb}$ is obtained as $U_{lb} = \mathcal{U} \mathcal{V}^{\dagger}$.
After all $U_{l b}$ in layer $l$ are updated for a predefined number of iterations,
the optimization moves to the next layer.
This sweeping process is repeated until convergence.

Block2 library~\cite{10.1063/5.0180424} is used to
calculate the ground state and
the first excited state of benzene via state-averaged,
electron-number-conserving DMRG with maximum bond dimension
$\chi = 256$. The electron numbers in each spin sector are not preserved.
An interleaved spin ordering
(i.e., $1\uparrow, 1 \downarrow, 2 \uparrow, 2 \downarrow, \dots$)
is employed without reordering.
The compression of the DMRG wavefunctions is done with
a maximum bond dimension $\chi' = 256$ for $\mathcal{L}^{l}$
and $\mathcal{R}^{l}$ and a maximum of 500 sweeps.

\begin{figure}
\includegraphics{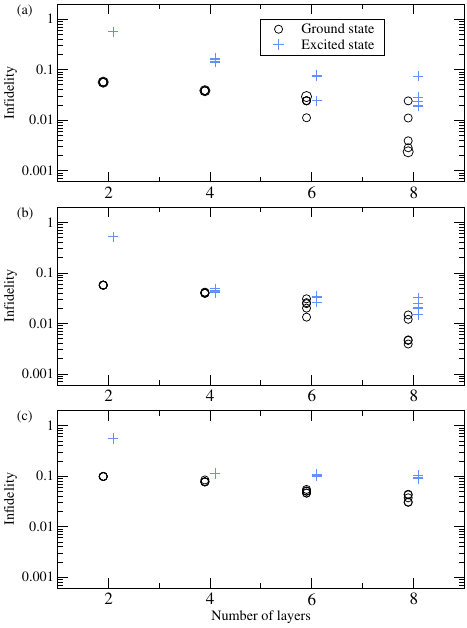}%
\caption{Infidelity between the original DMRG 
and the approximated wavefunctions for (a) 16, (b) 24, and (c) 32
qubit states.}
\label{fig:infidelity_all}
\end{figure}

Figure~\ref{fig:infidelity_all} shows the calculated 
infidelity $1 - |\langle \Psi_{\textrm{DMRG}}|\tilde{\Psi}\rangle|^{2}$
of the ground state and the first excited state for the three cases
as a function of the number of layers $L$.
In each case five different random initial states are used.
In all cases the infidelity is below $\approx 0.05$ for
the ground state and $\approx 0.1$ for the excited state
with $L \geq 6$. We therefore use the $L=6$ results with the lowest infidelity
in our hardware demonstration.

Each two-qubit unitary $U_{l b}$ is
converted into quantum gates with up to three two-qubit gates.
We apply \texttt{qiskit.synthesis.two\_qubit\_decompose} function in Qiskit, which
internally uses the KAK decomposition~\cite{PhysRevA.100.032328}.
For the ground state (the excited state),
the quantum circuits transpiled by Qiskit
contain 98, 153, and 227 (110, 174, and 227) two-qubit basis gates (\texttt{cz} gates)
for 16, 24, and 32 qubits, respectively.

\subsection{Error mitigation}
\begin{figure*}[ht]
\includegraphics{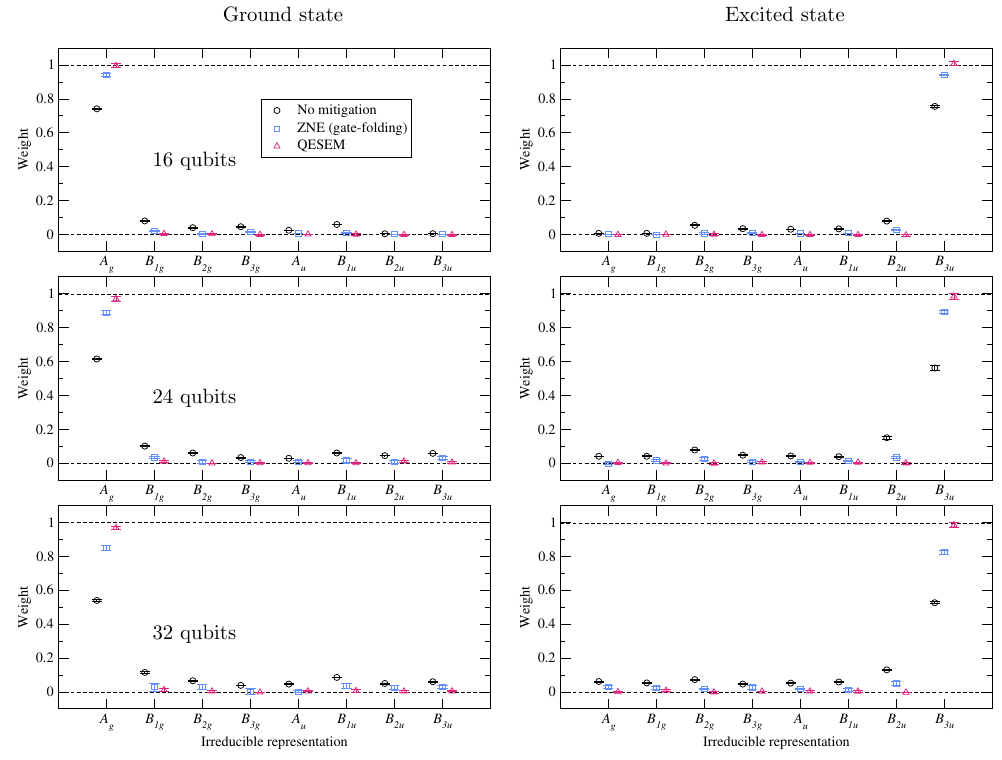}%
\caption{Estimated weights $w_{\Gamma}$ for the ground state and the first excited state of
benzene in $D_{2h}$ symmetry on \texttt{ibm\_kawasaki} with 16, 24, and 32 qubits.
Each data point is an average of five trials, with error bars indicating the standard error.
The two horizontal dashed lines in each panel
represent zero and the exact weight (see Table~\ref{tbl:weights_kawasaki}).
}
\label{fig:weights_all}
\end{figure*}
\begin{table*}[ht]
  \centering
  \begin{tabular}{lllll}\hline
      &                                           & (10e,8o) & (10e,12o) & (10e,16o) \\\hline
  No mitigation          & Ground  state          & $0.742 \pm 0.005$ & $0.615 \pm 0.003$ & $0.540 \pm 0.007$ \\
                         & Excited state          & $0.756 \pm 0.005$ & $0.562 \pm 0.015$ & $0.526 \pm 0.007$ \\
  ZNE (gate-folding)     & Ground  state          & $0.940 \pm 0.007$ & $0.889 \pm 0.011$ & $0.849 \pm 0.014$ \\
                         & Excited state          & $0.941 \pm 0.004$ & $0.892 \pm 0.010$ & $0.824 \pm 0.013$ \\
  QESEM                  & Ground  state          & $0.996 \pm 0.011$ & $0.969 \pm 0.015$ & $0.968 \pm 0.010$ \\
                         & Excited state          & $1.008 \pm 0.013$ & $0.985 \pm 0.017$ & $0.985 \pm 0.013$ \\
  Exact                  & Ground  state          & $0.998$ & $0.997$ & $0.999$ \\
                         & Excited state          & $1.000$ & $0.996$ & $0.994$ \\ \hline
  \end{tabular}
  \caption{Estimated weights of $A_{g}$ for the ground state and $B_{3u}$ for the excited state of benzene.
  \label{tbl:weights_kawasaki}}
\end{table*}
We apply two error mitigation techniques:
(i) gate-folding based zero noise extrapolation (ZNE) implemented in Qiskit
and
(ii) quasi-probabilistic error mitigation approach implemented in 
QESEM~\cite{aharonov2025reliablehighaccuracyerrormitigation}
available via Qiskit Function~\cite{QESEMQiskitFunction}.

The gate-folding based ZNE amplifies noise by replacing specific gates (two-qubit gates)
with equivalent but redundant gate sequences (e.g. $U \to U U^{\dagger} U$).
The obtained noisy results are extrapolated to get noise-free
expectation value results. The extrapolated results
are not guaranteed to be unbiased.
We set Qiskit option \texttt{resilience\_level}=2~\cite{ConfigureErrorMitigation}, which
uses ZNE noise factors $(1, 3, 5)$. This option
also activates readout error mitigation and measurement twirling
via the model-free technique called twirled readout error extinction (TREX)~\cite{PhysRevA.105.032620}.

Unlike ZNE, the quasi-probability error mitigation method underlying QESEM is an unbiased method.
QESEM extends the applicability of probabilistic error cancellation (PEC)~\cite{PhysRevLett.119.180509,vandenBerg2023}
by incorporating techniques such as the active-volume identification and 
multi-type quasi-probability decompositions which allow mitigation of non-Clifford two-qubit gates~\cite{aharonov2025reliablehighaccuracyerrormitigation}.
QESEM first performs device characterization, and this information is used for
error suppression, noise-aware transpilation, error model construction, and building quasi-probability decomposition for implementing the inverse noise channels.
The mitigated results and their variances are obtained by processing the measurement data
of mitigation circuits sampled from the ensemble defined by the quasi-probability decomposition.
More details of this approach are described in Ref.~\cite{aharonov2025reliablehighaccuracyerrormitigation}.

For each setup, we carry out five independent sets of measurements with a target precision of 0.02.
The corresponding number of shots per circuit in ZNE is about 2500.
In QESEM, the number of total shots for each measurement set, including calibration,
noise characterization, and mitigation, with 32 qubits
varies from $ 1.4 \times 10^6$ to $2.1 \times 10^6$.
The corresponding QPU time for each measurement set ranges approximately from 400 to 600 seconds.

\subsection{Hardware results}

Figure~\ref{fig:weights_all}
shows the hardware results executed
on \texttt{ibm\_kawasaki} for the ground state and the first excited state.
The correct irreducible representations are $A_{g}$ and $B_{3u}$ for the ground state
and the excited state, respectively.
The estimated weights of these irreducible
representations are summarized in Table~\ref{tbl:weights_kawasaki}.
Note that the exact results in Table~\ref{tbl:weights_kawasaki}
deviate slightly from unity due to the approximate encoding.

Without error mitigation, the weights of the correct representations decrease as the number of qubits increases.
In the 32-qubit results, the weights reduce to 0.540 for the ground state and 0.526 for the excited state.
To further assess the quality of raw measurement results,
independent measurements are performed
which count the number of bitstrings with correct electron numbers, using $10^{5}$ shots each.
The probabilities of obtaining correct bitstrings for the ground state
are 0.70, 0.55, and 0.36 for 16, 24, and 32 qubits, respectively.
The corresponding values for the excited state are 0.68, 0.46, and 0.35, respectively.
These poor results clearly show the importance of error mitigation.

The gate-folding based ZNE improves the quality of the results, but the results are biased.
The results with QESEM's quasi-probabilistic approach show the best agreement,
with a maximum discrepancy of only a few percent.
This comparison highlights the importance of accurate noise characterization and removal,
and also validates the efficacy of our proposed method.
This demonstration also suggests that
our method can be used for benchmarking quantum devices
and error mitigation techniques,
as our method requires no ancilla qubit and no or very little overhead for symmetry-adapted orbitals.

Although the present demonstration is based on the Pauli-based evaluation, which works only for abelian groups,
our proposed method can also be applied to non-abelian symmetry groups with arbitrary basis functions.
The hardware application of our method to a non-abelian case is an interesting direction for future research.

\section{Conclusion}
\label{sec:conclusion}
In this work, we presented a practical framework for studying point-group symmetry properties of many-electron wavefunctions
on a quantum computer.
We proposed an ancilla-free, basis-agnostic method to evaluate the weights of irreducible representations in a given
wavefunction, which works for both abelian and non-abelian groups.
Our numerical simulations on single Slater determinants and correlated wavefunctions
show the usefulness of our method in preparing a trial state with a specific symmetry,
as well as in assessing the quality of a wavefunction.

We also presented a hardware demonstration of our proposed method using up to 32 qubits of
IBM's superconducting quantum device \texttt{ibm\_kawasaki}.
By combining a tensor-network based state preparation scheme and advanced error mitigation techniques,
we successfully reproduced the correct
weights of the irreducible representations for the ground state and the excited state of
benzene.

We expect our methodology to be useful for analyzing many-electron states prepared with more advanced quantum algorithms,
and to serve as a tool for benchmarking quantum devices and error mitigation techniques.
The extension of the proposed method to other symmetry groups is also an interesting topic for future study.

\begin{acknowledgments}
R.S. thanks Ori Alberton, Asaf Berkovitch, Netanel Lindner, and Asif Sinay
for technical advice on the use of QESEM.
This work is partly supported by UTokyo Quantum Initiative.
Molecular structures were visualized using VESTA~\cite{Momma:db5098}.
\end{acknowledgments}

\section*{Author Contributions}
R.S. and H.N. conceived the idea and developed the methodology.
R.S. developed calculation codes, performed numerical simulations and hardware demonstrations, and edited the manuscript.
K.S. contributed to the selection of target molecules and the interpretation of the results.
S.K. provided advice on the development of the tensor-network based method.
T.I. provided guidance on quantum circuit construction and the use of quantum hardware.
All authors discussed the results and reviewed and approved the final manuscript.
Large language models were used for grammar checking and improving the phrasing of the manuscript.

\appendix*
\section{Derivation of $\sum_{\Gamma} w_{\Gamma} = 1$}
\label{sec:sum_w_gamma}
Here we show $\sum_{\Gamma} w_{\Gamma} = 1$ is satisfied for $w_{\Gamma}$ given in Eq.~(\ref{eq:w_gamma}).
The derivation is based on the orthogonality relationships of $\chi_{\Gamma}(C)$~\cite{bishop}
\begin{eqnarray}
  \frac{1}{|G|}\sum_{C} r_{C} \chi^{*}_{\Gamma}(C) \chi_{\Gamma'}(C) &=&  \delta_{\Gamma \Gamma'}
  \label{eq:chiorth1}\\
  \frac{1}{|G|}\sum_{\Gamma} \chi^{*}_{\Gamma}(C) \chi_{\Gamma}(C') &=& \frac{1}{r_{C}} \delta_{C C'}
  \label{eq:chiorth2}
\end{eqnarray}
where 
\begin{equation}
  r_{C} = \sum_{g \in C} 1
  \label{eq:r_c}
\end{equation}
is the number of elements in $C$.
Noting that
\begin{equation}
  d_{\Gamma} = \chi_{\Gamma}(E)
\end{equation}
with $E$ the identity operation, from Eq.~(\ref{eq:w_gamma}) 
\begin{eqnarray}
  \sum_{\Gamma} w_{\Gamma} &=&
  \sum_{C}\frac{1}{|G|}
  \sum_{\Gamma}
  \chi_{\Gamma}^{*}(C) \chi_{\Gamma}(E)
  \sum_{g \in C} \langle \Psi |\hat{g}|\Psi\rangle \nonumber\\
  &=&
  \sum_{C} \frac{\delta_{C E}}{r_{E}}
  \sum_{g \in C} \langle \Psi |\hat{g}|\Psi\rangle \nonumber\\
  &=& \langle \Psi |\hat{E}|\Psi\rangle =1.
  \label{eq:sum_w_gamma}
\end{eqnarray}
Here $r_{E} = 1$ is used.
Equation~(\ref{eq:sum_w_gamma}) shows that $\sum_{\Gamma} w_{\Gamma} = 1$ is satisfied independently of $\langle \Psi | \hat{g}|\Psi \rangle$ for $g \neq E$.
\bibliography{refs}

@article{PhysRevB.83.115125,
  title = {Tensor network states and algorithms in the presence of a global {U(1)} symmetry},
  author = {Singh, Sukhwinder and Pfeifer, Robert N. C. and Vidal, Guifre},
  journal = {Phys. Rev. B},
  volume = {83},
  issue = {11},
  pages = {115125},
  numpages = {22},
  year = {2011},
  month = {Mar},
  publisher = {American Physical Society},
  doi = {10.1103/PhysRevB.83.115125},
  url = {https://link.aps.org/doi/10.1103/PhysRevB.83.115125}
}

@article{Gard2020,
  author = {Gard, Bryan T. and Zhu, Linghua and Barron, George S. and Mayhall, Nicholas J. and Economou, Sophia E. and Barnes, Edwin},
  pages = {10},
  title = {Efficient symmetry-preserving state preparation circuits for the variational quantum eigensolver algorithm},
  journal = {npj Quantum Information},
  year = {2020},
  volume = {6},
  number = {1},
  doi = {10.1038/s41534-019-0240-1},
  url = {https://doi.org/10.1038/s41534-019-0240-1},
  issn = {2056-6387},
}

@misc{bravyi2017taperingqubitssimulatefermionic,
      title={Tapering off qubits to simulate fermionic {H}amiltonians}, 
      author={Sergey Bravyi and Jay M. Gambetta and Antonio Mezzacapo and Kristan Temme},
      year={2017},
      eprint={1701.08213},
      archivePrefix={arXiv},
      primaryClass={quant-ph},
      url={https://arxiv.org/abs/1701.08213}, 
}

@book{bishop,
  title={Group Theory and Chemistry},
  author={Bishop, D.M.},
  isbn={9780486132327},
  series={Dover Books on Chemistry},
  year={2012},
  publisher={Dover Publications}
}

@article{10.1098/rspa.1937.0142,
    author = {Jahn, H. A. and Teller, E.},
    title = {{Stability of polyatomic molecules in degenerate electronic states - I—Orbital degeneracy}},
    journal = {Proceedings of the Royal Society of London. A. Mathematical and Physical Sciences},
    volume = {161},
    number = {905},
    pages = {220-235},
    year = {1937},
    month = {07},
    issn = {0080-4630},
    doi = {10.1098/rspa.1937.0142},
    url = {https://doi.org/10.1098/rspa.1937.0142}
}

@article{10.1098/rspa.1938.0008,
    author = {Jahn, H. A.},
    title = {{Stability of polyatomic molecules in degenerate electronic states II-Spin degeneracy}},
    journal = {Proceedings of the Royal Society of London. A. Mathematical and Physical Sciences},
    volume = {164},
    number = {916},
    pages = {117-131},
    year = {1938},
    month = {01},
    issn = {0080-4630},
    doi = {10.1098/rspa.1938.0008},
    url = {https://doi.org/10.1098/rspa.1938.0008}
}

@book{bersuker,
  title={The Jahn-Teller Effect},
  author={Bersuker, I.B.},
  isbn={9780521822121},
  year={2006},
  publisher={Cambridge University Press}
}

@article{Woodward1965,
  author = {Woodward, R. B. and Hoffmann, Roald},
  pages = {395--397},
  title = {Stereochemistry of Electrocyclic Reactions},
  journal = {J. Am. Chem. Soc.},
  year = {1965},
  volume = {87},
  number = {2},
  url = {https://doi.org/10.1021/ja01080a054},
  publisher = {American Chemical Society},
  issn = {0002-7863}
}

@article{10.1063/1.5110682,
    author = {Yen, Tzu-Ching and Lang, Robert A. and Izmaylov, Artur F.},
    title = {Exact and approximate symmetry projectors for the electronic structure problem on a quantum computer},
    journal = {The Journal of Chemical Physics},
    volume = {151},
    number = {16},
    pages = {164111},
    year = {2019},
    month = {10},
    issn = {0021-9606},
    doi = {10.1063/1.5110682},
    url = {https://doi.org/10.1063/1.5110682}
}

@Article{Setia2020,
author={Setia, Kanav
and Chen, Richard
and Rice, Julia E.
and Mezzacapo, Antonio
and Pistoia, Marco
and Whitfield, James D.},
title={Reducing Qubit Requirements for Quantum Simulations Using Molecular Point Group Symmetries},
journal={Journal of Chemical Theory and Computation},
year={2020},
month={Oct},
day={13},
publisher={American Chemical Society},
volume={16},
number={10},
pages={6091-6097},
issn={1549-9618},
doi={10.1021/acs.jctc.0c00113},
url={https://doi.org/10.1021/acs.jctc.0c00113}
}

@article{Picozzi_2023,
doi = {10.1088/2058-9565/acd86c},
url = {https://doi.org/10.1088/2058-9565/acd86c},
year = {2023},
month = {jun},
publisher = {IOP Publishing},
volume = {8},
number = {3},
pages = {035026},
author = {Picozzi, Dario and Tennyson, Jonathan},
title = {Symmetry-adapted encodings for qubit number reduction by point-group and other {B}oolean symmetries},
journal = {Quantum Science and Technology}
}

@article{PhysRevA.105.062452,
  title = {Progress toward larger molecular simulation on a quantum computer: {S}imulating a system with up to 28 qubits accelerated by point-group symmetry},
  author = {Cao, Changsu and Hu, Jiaqi and Zhang, Wengang and Xu, Xusheng and Chen, Dechin and Yu, Fan and Li, Jun and Hu, Han-Shi and Lv, Dingshun and Yung, Man-Hong},
  journal = {Phys. Rev. A},
  volume = {105},
  issue = {6},
  pages = {062452},
  numpages = {12},
  year = {2022},
  month = {Jun},
  publisher = {American Physical Society},
  doi = {10.1103/PhysRevA.105.062452},
  url = {https://link.aps.org/doi/10.1103/PhysRevA.105.062452}
}

@article{PhysRevA.111.052433,
  title = {Unification of finite symmetries in the simulation of many-body systems on quantum computers},
  author = {Bastidas, Victor M. and Fitzpatrick, Nathan and Joven, K. J. and Rossi, Zane M. and Islam, Shariful and Van Voorhis, Troy and Chuang, Isaac L. and Liu, Yuan},
  journal = {Phys. Rev. A},
  volume = {111},
  issue = {5},
  pages = {052433},
  numpages = {27},
  year = {2025},
  month = {May},
  publisher = {American Physical Society},
  doi = {10.1103/PhysRevA.111.052433},
  url = {https://link.aps.org/doi/10.1103/PhysRevA.111.052433}
}

@article{10.1145/1198513.1198525,
author = {Moore, Cristopher and Rockmore, Daniel and Russell, Alexander},
title = {Generic quantum Fourier transforms},
year = {2006},
issue_date = {October 2006},
publisher = {Association for Computing Machinery},
address = {New York, NY, USA},
volume = {2},
number = {4},
issn = {1549-6325},
url = {https://doi.org/10.1145/1198513.1198525},
doi = {10.1145/1198513.1198525},
journal = {ACM Trans. Algorithms},
month = oct,
pages = {707–723},
numpages = {17}
}

@misc{khinevich2026symmetryadaptedstatepreparationquantum,
      title={Symmetry-Adapted State Preparation for Quantum Chemistry on Fault-Tolerant Quantum Computers}, 
      author={Viktor Khinevich and Wataru Mizukami},
      year={2026},
      eprint={2601.08533},
      archivePrefix={arXiv},
      primaryClass={quant-ph},
      url={https://arxiv.org/abs/2601.08533}, 
}

@article{doi:10.1126/science.abb9811,
author = {Frank Arute  and Kunal Arya  and Ryan Babbush  and Dave Bacon  and Joseph C. Bardin  and Rami Barends  and Sergio Boixo  and Michael Broughton  and Bob B. Buckley  and David A. Buell  and Brian Burkett  and Nicholas Bushnell  and Yu Chen  and Zijun Chen  and Benjamin Chiaro  and Roberto Collins  and William Courtney  and Sean Demura  and Andrew Dunsworth  and Edward Farhi  and Austin Fowler  and Brooks Foxen  and Craig Gidney  and Marissa Giustina  and Rob Graff  and Steve Habegger  and Matthew P. Harrigan  and Alan Ho  and Sabrina Hong  and Trent Huang  and William J. Huggins  and Lev Ioffe  and Sergei V. Isakov  and Evan Jeffrey  and Zhang Jiang  and Cody Jones  and Dvir Kafri  and Kostyantyn Kechedzhi  and Julian Kelly  and Seon Kim  and Paul V. Klimov  and Alexander Korotkov  and Fedor Kostritsa  and David Landhuis  and Pavel Laptev  and Mike Lindmark  and Erik Lucero  and Orion Martin  and John M. Martinis  and Jarrod R. McClean  and Matt McEwen  and Anthony Megrant  and Xiao Mi  and Masoud Mohseni  and Wojciech Mruczkiewicz  and Josh Mutus  and Ofer Naaman  and Matthew Neeley  and Charles Neill  and Hartmut Neven  and Murphy Yuezhen Niu  and Thomas E. O’Brien  and Eric Ostby  and Andre Petukhov  and Harald Putterman  and Chris Quintana  and Pedram Roushan  and Nicholas C. Rubin  and Daniel Sank  and Kevin J. Satzinger  and Vadim Smelyanskiy  and Doug Strain  and Kevin J. Sung  and Marco Szalay  and Tyler Y. Takeshita  and Amit Vainsencher  and Theodore White  and Nathan Wiebe  and Z. Jamie Yao  and Ping Yeh  and Adam Zalcman },
title = {Hartree-{F}ock on a superconducting qubit quantum computer},
journal = {Science},
volume = {369},
number = {6507},
pages = {1084-1089},
year = {2020},
doi = {10.1126/science.abb9811},
URL = {https://www.science.org/doi/abs/10.1126/science.abb9811}
}

@article{Lacroix2023,
  author = {Lacroix, Denis and Ruiz Guzman, Edgar Andres and Siwach, Pooja},
  pages = {3},
  title = {Symmetry breaking/symmetry preserving circuits and symmetry restoration on quantum computers},
  journal = {The European Physical Journal A},
  year = {2023},
  volume = {59},
  number = {1},
  doi = {10.1140/epja/s10050-022-00911-7},
  url = {https://doi.org/10.1140/epja/s10050-022-00911-7},
  issn = {1434-601X},
}

@article{Chakraborty2024implementingany,
  doi = {10.22331/q-2024-10-10-1496},
  url = {https://doi.org/10.22331/q-2024-10-10-1496},
  title = {Implementing any {L}inear {C}ombination of {U}nitaries on {I}ntermediate-term {Q}uantum {C}omputers},
  author = {Chakraborty, Shantanav},
  journal = {{Quantum}},
  issn = {2521-327X},
  publisher = {{Verein zur F{\"{o}}rderung des Open Access Publizierens in den Quantenwissenschaften}},
  volume = {8},
  pages = {1496},
  month = oct,
  year = {2024}
}

@misc{sung2026ffsimfastersimulationfermionic,
      title={ffsim: Faster simulation of fermionic quantum circuits}, 
      author={Kevin J. Sung and Inho Choi and Mirko Amico and Bartholomew Andrews and Esra Ayantuna and Yukio Kawashima and Wan-Hsuan Lin and David Omanovic and Samuele Piccinelli and Javier Robledo Moreno and Abdullah Ash Saki and James Shee and Soyoung Shin and Minh C. Tran and Kento Ueda and Haimeng Zhang and Mario Motta},
      year={2026},
      eprint={2605.03123},
      archivePrefix={arXiv},
      primaryClass={quant-ph},
      url={https://arxiv.org/abs/2605.03123}, 
}

@misc{ffsim,
  author = {{The ffsim developers}},
  title = {ffsim: {F}aster simulations of fermionic quantum circuits},
  howpublished = {\url{https://github.com/qiskit-community/ffsim}}
}

@Article{D3SC02516K,
author ="Motta, Mario and Sung, Kevin J. and Whaley, K. Birgitta and Head-Gordon, Martin and Shee, James",
title  ={Bridging physical intuition and hardware efficiency for correlated electronic states: the local unitary cluster {J}astrow ansatz for electronic structure},
journal  ="Chem. Sci.",
year  ="2023",
volume  ="14",
issue  ="40",
pages  ="11213-11227",
publisher  ="The Royal Society of Chemistry",
doi  ="10.1039/D3SC02516K",
url  ="http://dx.doi.org/10.1039/D3SC02516K"
}

@article{
doi:10.1126/sciadv.adu9991,
author = {Javier Robledo-Moreno  and Mario Motta  and Holger Haas  and Ali Javadi-Abhari  and Petar Jurcevic  and William Kirby  and Simon Martiel  and Kunal Sharma  and Sandeep Sharma  and Tomonori Shirakawa  and Iskandar Sitdikov  and Rong-Yang Sun  and Kevin J. Sung  and Maika Takita  and Minh C. Tran  and Seiji Yunoki  and Antonio Mezzacapo },
title = {Chemistry beyond the scale of exact diagonalization on a quantum-centric supercomputer},
journal = {Science Advances},
volume = {11},
number = {25},
pages = {eadu9991},
year = {2025},
doi = {10.1126/sciadv.adu9991},
URL = {https://www.science.org/doi/abs/10.1126/sciadv.adu9991}
}

@misc{lin2025improvedparameterinitializationlocal,
      title={Improved parameter initialization for the (local) unitary cluster {J}astrow ansatz}, 
      author={Wan-Hsuan Lin and Fangchun Liang and Mario Motta and Haimeng Zhang and Kenneth M. Merz Jr. and Kevin J. Sung},
      year={2025},
      eprint={2511.22476},
      archivePrefix={arXiv},
      primaryClass={quant-ph},
      url={https://arxiv.org/abs/2511.22476}, 
}

@article{doi:10.1021/ed048p92,
author = {Ellis, R. L. and Jaffe, H. H.},
title = {The symmetries and multiplicities of electronic states in polyatomic molecules},
journal = {Journal of Chemical Education},
volume = {48},
number = {2},
pages = {92},
year = {1971},
doi = {10.1021/ed048p92},
URL = { 
        https://doi.org/10.1021/ed048p92
}
}

@article{MOHAMMADI201251,
title = {Differentiation of ferrocene {D5d} and {D5h} conformers using {IR} spectroscopy},
journal = {Journal of Organometallic Chemistry},
volume = {713},
pages = {51-59},
year = {2012},
issn = {0022-328X},
doi = {https://doi.org/10.1016/j.jorganchem.2012.04.009},
url = {https://www.sciencedirect.com/science/article/pii/S0022328X12002082},
author = {Narges Mohammadi and Aravindhan Ganesan and Christopher T. Chantler and Feng Wang}
}

@article{PhysRevResearch.6.043008,
  title = {Automatic quantum circuit encoding of a given arbitrary quantum state},
  author = {Shirakawa, Tomonori and Ueda, Hiroshi and Yunoki, Seiji},
  journal = {Phys. Rev. Res.},
  volume = {6},
  issue = {4},
  pages = {043008},
  numpages = {26},
  year = {2024},
  month = {Oct},
  publisher = {American Physical Society},
  doi = {10.1103/PhysRevResearch.6.043008},
  url = {https://link.aps.org/doi/10.1103/PhysRevResearch.6.043008}
}

@article{
doi:10.1073/pnas.2425026122,
author = {Shu Kanno  and Kenji Sugisaki  and Hajime Nakamura  and Hiroshi Yamauchi  and Rei Sakuma  and Takao Kobayashi  and Qi Gao  and Naoki Yamamoto},
title = {Tensor-based quantum phase difference estimation for large-scale demonstration},
journal = {Proceedings of the National Academy of Sciences},
volume = {122},
number = {30},
pages = {e2425026122},
year = {2025},
doi = {10.1073/pnas.2425026122},
URL = {https://www.pnas.org/doi/abs/10.1073/pnas.2425026122}
}

@article{10.1063/5.0180424,
    author = {Zhai, Huanchen and Larsson, Henrik R. and Lee, Seunghoon and Cui, Zhi-Hao and Zhu, Tianyu and Sun, Chong and Peng, Linqing and Peng, Ruojing and Liao, Ke and Tölle, Johannes and Yang, Junjie and Li, Shuoxue and Chan, Garnet Kin-Lic},
    title = {Block2: {A} comprehensive open source framework to develop and apply state-of-the-art {DMRG} algorithms in electronic structure and beyond},
    journal = {The Journal of Chemical Physics},
    volume = {159},
    number = {23},
    pages = {234801},
    year = {2023},
    month = {12},
    issn = {0021-9606},
    doi = {10.1063/5.0180424},
    url = {https://doi.org/10.1063/5.0180424}
}

@article{10.1063/5.0006074,
    author = {Sun, Qiming and Zhang, Xing and Banerjee, Samragni and Bao, Peng and Barbry, Marc and Blunt, Nick S. and Bogdanov, Nikolay A. and Booth, George H. and Chen, Jia and Cui, Zhi-Hao and Eriksen, Janus J. and Gao, Yang and Guo, Sheng and Hermann, Jan and Hermes, Matthew R. and Koh, Kevin and Koval, Peter and Lehtola, Susi and Li, Zhendong and Liu, Junzi and Mardirossian, Narbe and McClain, James D. and Motta, Mario and Mussard, Bastien and Pham, Hung Q. and Pulkin, Artem and Purwanto, Wirawan and Robinson, Paul J. and Ronca, Enrico and Sayfutyarova, Elvira R. and Scheurer, Maximilian and Schurkus, Henry F. and Smith, James E. T. and Sun, Chong and Sun, Shi-Ning and Upadhyay, Shiv and Wagner, Lucas K. and Wang, Xiao and White, Alec and Whitfield, James Daniel and Williamson, Mark J. and Wouters, Sebastian and Yang, Jun and Yu, Jason M. and Zhu, Tianyu and Berkelbach, Timothy C. and Sharma, Sandeep and Sokolov, Alexander Yu. and Chan, Garnet Kin-Lic},
    title = {Recent developments in the {PySCF} program package},
    journal = {The Journal of Chemical Physics},
    volume = {153},
    number = {2},
    pages = {024109},
    year = {2020},
    month = {07},
    issn = {0021-9606},
    doi = {10.1063/5.0006074},
    url = {https://doi.org/10.1063/5.0006074}
}

@article{PhysRevLett.69.2863,
  title = {Density matrix formulation for quantum renormalization groups},
  author = {White, Steven R.},
  journal = {Phys. Rev. Lett.},
  volume = {69},
  issue = {19},
  pages = {2863--2866},
  numpages = {0},
  year = {1992},
  month = {Nov},
  publisher = {American Physical Society},
  doi = {10.1103/PhysRevLett.69.2863},
  url = {https://link.aps.org/doi/10.1103/PhysRevLett.69.2863}
}

@article{SCHOLLWOCK201196,
title = {The density-matrix renormalization group in the age of matrix product states},
journal = {Annals of Physics},
volume = {326},
number = {1},
pages = {96-192},
year = {2011},
note = {January 2011 Special Issue},
issn = {0003-4916},
doi = {https://doi.org/10.1016/j.aop.2010.09.012},
url = {https://www.sciencedirect.com/science/article/pii/S0003491610001752},
author = {Ulrich Schollwöck}
}

@article{PhysRevX.10.041038,
  title = {What Limits the Simulation of Quantum Computers?},
  author = {Zhou, Yiqing and Stoudenmire, E. Miles and Waintal, Xavier},
  journal = {Phys. Rev. X},
  volume = {10},
  issue = {4},
  pages = {041038},
  numpages = {15},
  year = {2020},
  month = {Nov},
  publisher = {American Physical Society},
  doi = {10.1103/PhysRevX.10.041038},
  url = {https://link.aps.org/doi/10.1103/PhysRevX.10.041038}
}

@misc{javadiabhari2024quantumcomputingqiskit,
      title={Quantum computing with Qiskit}, 
      author={Ali Javadi-Abhari and Matthew Treinish and Kevin Krsulich and Christopher J. Wood and Jake Lishman and Julien Gacon and Simon Martiel and Paul D. Nation and Lev S. Bishop and Andrew W. Cross and Blake R. Johnson and Jay M. Gambetta},
      year={2024},
      eprint={2405.08810},
      archivePrefix={arXiv},
      primaryClass={quant-ph},
      url={https://arxiv.org/abs/2405.08810}, 
}

@book{dreizler2012density,
  title={Density Functional Theory: An Approach to the Quantum Many-Body Problem},
  author={Dreizler, R.M. and Gross, E.K.U.},
  year={2012},
  publisher={Springer Berlin Heidelberg}
}

@book{doi:https://doi.org/10.1002/9781119019572,
title = {Molecular Electronic‐Structure Theory},
author={Helgaker, Trygve and J{\o}rgensen, Poul and Olsen, Jeppe},
year = {2000},
doi = {https://doi.org/10.1002/9781119019572},
publisher = {John Wiley \& Sons, Ltd}
}

@article{Rubin_2018,
doi = {10.1088/1367-2630/aab919},
url = {https://doi.org/10.1088/1367-2630/aab919},
year = {2018},
month = {may},
publisher = {IOP Publishing},
volume = {20},
number = {5},
pages = {053020},
author = {Rubin, Nicholas C and Babbush, Ryan and McClean, Jarrod},
title = {Application of fermionic marginal constraints to hybrid quantum algorithms},
journal = {New Journal of Physics}
}

@article{PRXQuantum.3.040305,
  title = {Ground-State Preparation and Energy Estimation on Early Fault-Tolerant Quantum Computers via Quantum Eigenvalue Transformation of Unitary Matrices},
  author = {Dong, Yulong and Lin, Lin and Tong, Yu},
  journal = {PRX Quantum},
  volume = {3},
  issue = {4},
  pages = {040305},
  numpages = {25},
  year = {2022},
  month = {Oct},
  publisher = {American Physical Society},
  doi = {10.1103/PRXQuantum.3.040305},
  url = {https://link.aps.org/doi/10.1103/PRXQuantum.3.040305}
}

@misc{kanno2023quantumselectedconfigurationinteractionclassical,
      title={{Quantum-Selected Configuration Interaction: classical diagonalization of Hamiltonians in subspaces selected by quantum computers}}, 
      author={Keita Kanno and Masaya Kohda and Ryosuke Imai and Sho Koh and Kosuke Mitarai and Wataru Mizukami and Yuya O. Nakagawa},
      year={2023},
      eprint={2302.11320},
      archivePrefix={arXiv},
      primaryClass={quant-ph},
      url={https://arxiv.org/abs/2302.11320}, 
}

@article{PhysRevA.100.032328,
  title = {Validating quantum computers using randomized model circuits},
  author = {Cross, Andrew W. and Bishop, Lev S. and Sheldon, Sarah and Nation, Paul D. and Gambetta, Jay M.},
  journal = {Phys. Rev. A},
  volume = {100},
  issue = {3},
  pages = {032328},
  numpages = {11},
  year = {2019},
  month = {Sep},
  publisher = {American Physical Society},
  doi = {10.1103/PhysRevA.100.032328},
  url = {https://link.aps.org/doi/10.1103/PhysRevA.100.032328}
}

@misc{aharonov2025importanceerrormitigationquantum,
      title={On the Importance of Error Mitigation for Quantum Computation}, 
      author={Dorit Aharonov and Ori Alberton and Itai Arad and Yosi Atia and Eyal Bairey and Zvika Brakerski and Itsik Cohen and Omri Golan and Ilya Gurwich and Oded Kenneth and Eyal Leviatan and Netanel H. Lindner and Ron Aharon Melcer and Adiel Meyer and Gili Schul and Maor Shutman},
      year={2025},
      eprint={2503.17243},
      archivePrefix={arXiv},
      primaryClass={quant-ph},
      url={https://arxiv.org/abs/2503.17243}, 
}

@misc{aharonov2025reliablehighaccuracyerrormitigation,
      title={Reliable high-accuracy error mitigation for utility-scale quantum circuits}, 
      author={Dorit Aharonov and Ori Alberton and Itai Arad and Yosi Atia and Eyal Bairey and Matan Ben Dov and Asaf Berkovitch and Zvika Brakerski and Itsik Cohen and Eran Fuchs and Omri Golan and Or Golan and Barak D. Gur and Ilya Gurwich and Avieli Haber and Rotem Haber and Dorri Halbertal and Yaron Itkin and Barak A. Katzir and Oded Kenneth and Shlomi Kotler and Roei Levi and Eyal Leviatan and Yotam Y. Lifshitz and Adi Ludmer and Shlomi Matityahu and Ron Aharon Melcer and Adiel Meyer and Omrie Ovdat and Aviad Panahi and Gil Ron and Ittai Rubinstein and Gili Schul and Tali Shnaider and Maor Shutman and Asif Sinay and Tasneem Watad and Assaf Zubida and Netanel H. Lindner},
      year={2025},
      eprint={2508.10997},
      archivePrefix={arXiv},
      primaryClass={quant-ph},
      url={https://arxiv.org/abs/2508.10997}, 
}

@article{PhysRevLett.119.180509,
  title = {Error Mitigation for Short-Depth Quantum Circuits},
  author = {Temme, Kristan and Bravyi, Sergey and Gambetta, Jay M.},
  journal = {Phys. Rev. Lett.},
  volume = {119},
  issue = {18},
  pages = {180509},
  numpages = {5},
  year = {2017},
  month = {Nov},
  publisher = {American Physical Society},
  doi = {10.1103/PhysRevLett.119.180509},
  url = {https://link.aps.org/doi/10.1103/PhysRevLett.119.180509}
}

@misc{QESEMQiskitFunction,
  title = {{QESEM: A Qiskit Function by Qedma}},
  year = {2026},
  howpublished = {\url{https://quantum.cloud.ibm.com/docs/en/guides/qedma-qesem}},
  note = {Accessed: 2026-04-15}
}

@article{PhysRevA.105.032620,
  title = {Model-free readout-error mitigation for quantum expectation values},
  author = {van den Berg, Ewout and Minev, Zlatko K. and Temme, Kristan},
  journal = {Phys. Rev. A},
  volume = {105},
  issue = {3},
  pages = {032620},
  numpages = {8},
  year = {2022},
  month = {Mar},
  publisher = {American Physical Society},
  doi = {10.1103/PhysRevA.105.032620},
  url = {https://link.aps.org/doi/10.1103/PhysRevA.105.032620}
}

@misc{ConfigureErrorMitigation,
  author = {{IBM Quantum}},
  title = {{Configure error mitigation}},
  year = {2026},
  howpublished = {\url{https://quantum.cloud.ibm.com/docs/en/guides/configure-error-mitigation}},
  note = {Accessed: 2026-04-15}
}

@Article{vandenBerg2023,
author={van den Berg, Ewout
and Minev, Zlatko K.
and Kandala, Abhinav
and Temme, Kristan},
title={Probabilistic error cancellation with sparse {P}auli--{L}indblad models on noisy quantum processors},
journal={Nature Physics},
year={2023},
month={Aug},
day={01},
volume={19},
number={8},
pages={1116-1121},
issn={1745-2481},
doi={10.1038/s41567-023-02042-2},
url={https://doi.org/10.1038/s41567-023-02042-2}
}

@article{Momma:db5098,
author = "Momma, Koichi and Izumi, Fujio",
title = "{{\it VESTA3} for three-dimensional visualization of crystal, volumetric and morphology data}",
journal = "Journal of Applied Crystallography",
year = "2011",
volume = "44",
number = "6",
pages = "1272--1276",
month = "Dec",
doi = {10.1107/S0021889811038970},
url = {https://doi.org/10.1107/S0021889811038970},
}

@article{Sohn-1971, 
    author = {Y. S. Sohn and D. N. Hendrickson and H. B. Gray}, 
    title = {Electronic structure of metallocene}, 
    journal = {J. Am. Chem. Soc.}, 
    year = {1971}, 
    volume = {93}, 
    number = {15}, 
    pages = {3603--3612}, 
    doi = {10.1021/ja00744a011}
}

@misc{kitaev1995quantummeasurementsabelianstabilizer,
      title={Quantum measurements and the Abelian Stabilizer Problem},
      author={A. Yu. Kitaev},
      year={1995},
      eprint={quant-ph/9511026},
      archivePrefix={arXiv},
      primaryClass={quant-ph},
      url={https://arxiv.org/abs/quant-ph/9511026},
}

\end{document}